%% file: pricing_paper.tex
\documentclass[12pt]{article}

\usepackage{booktabs}
\usepackage{siunitx}
\usepackage{dcolumn}
\usepackage{amsmath}
\usepackage{xfrac}
\usepackage{enumitem}
\usepackage[shortcuts]{extdash}
\usepackage[olditem,oldenum]{paralist}
\usepackage{amssymb}
\usepackage{setspace}

\usepackage{bbm}
\usepackage{float}
\usepackage{lscape}
\usepackage{array}
\usepackage{graphicx, color}
\usepackage[dvipsnames]{xcolor}
\usepackage{alltt}
\usepackage[T1]{fontenc}
\usepackage{amsmath}
\usepackage[top=1in, bottom=1in, left=1in, right=1in]{geometry}
\usepackage{sectsty}
\usepackage{amssymb}
\usepackage{placeins}
\usepackage{eurosym}
\usepackage{mathrsfs}
\usepackage{graphicx}
\usepackage{multirow}
\usepackage{MnSymbol}
\usepackage{systeme}
\usepackage[hidelinks]{hyperref}
\usepackage{fancyhdr}
\usepackage{alltt}
\usepackage{subfig}
\usepackage{rotating}
\usepackage{pgf}
\usepackage{dcolumn}
\usepackage{amssymb}
\usepackage{listings}
\usepackage{blindtext, xcolor}
\usepackage{booktabs,caption}
\usepackage[flushleft]{threeparttable}
\usepackage[title,toc,titletoc,page]{appendix}

\usepackage[sorting = nyt, style = apa, language = british, maxcitenames=3, mincitenames = 1,uniquename=false]{biblatex}
\usepackage{changepage}
\usepackage{ntheorem}
\usepackage[up,bf,raggedright]{titlesec}

\usepackage{etoc}

\usepackage{bm}
\usepackage{authblk}
\usepackage{array}

\usepackage{tikz}
\usetikzlibrary{shapes,arrows}
\usepackage{caption}
\usepackage{datenumber,xparse}
\usetikzlibrary{arrows.meta,backgrounds, calc,angles,positioning,intersections}

\newcounter{chronosstartdate}
\newcounter{chronosenddate}
\newcounter{chronosstartyear}
\newcounter{chronosendyear}
\newcounter{chronosyeardate}
\newcounter{chronosthingdate}
\newcounter{chronosotherthingdate}
\pgfkeys{/pgf/number format,
  int detect,
  set thousands separator={},
}
\tikzset{
  chronos/.code={
    \tikzset{%
      align=center,
      anchor=mid,
      /chronos/.cd,
      #1
    }%
    \setstartyear{\chronosstartyear}%
    \setmydatenumber{chronosstartdate}{\chronosstartyear}{\chronosstartmonth}{\chronosstartday}%
    \setmydatenumber{chronosenddate}{\chronosendyear}{\chronosendmonth}{\chronosendday}%
    \pgfmathsetmacro\chronosunit{(\chronoswidth-20pt)/(\thechronosenddate-\thechronosstartdate)}%
    \draw [line width=\chronosheight] (-10pt,0) coordinate (chronos pre) -- +(\chronoswidth,0) coordinate (chronos post);
    \coordinate (chronos start) at (0,0);
    \coordinate (chronos end) at ([xshift=-10pt]chronos post);
    \setcounter{chronosstartyear}{\chronosstartyear}%
    \setcounter{chronosendyear}{\chronosendyear}%
    \def\tempa{01}%
    \ifx\chronosstartmonth\tempa
      \ifx\chronosstartday\tempa
        \else\stepcounter{chronosstartyear}%
      \fi
      \else\stepcounter{chronosstartyear}%
    \fi
    \def\tempa{12}%
    \def\tempb{31}%
    \ifx\chronosendmonth\tempa
      \ifx\chronosendday\tempb
        \stepcounter{chronosendyear}%
      \fi
    \fi
    \foreach \i in {\thechronosstartyear,...,\thechronosendyear} {%
      \setmydatenumber{chronosyeardate}{\i}{01}{01}%
      \node [above, anchor=south, yshift=.5*\chronosheight] at ({(\thechronosyeardate-\thechronosstartdate)*\chronosunit pt},0) {\i}; }
  },
  chronos set date/.code args={#1-#2-#3:#4}{%
    \tikzset{%
      /chronos/.cd,
      #4 year={#1},
      #4 month={#2},
      #4 day={#3},
    }%
    \setmydatenumber{chronos#4date}{\csname chronos#4year\endcsname}{\csname chronos#4month\endcsname}{\csname chronos#4day\endcsname}%
  },
  chronos date/.style args={#1-#2-#3}{%
    chronos set date/.expanded={#1-#2-#3:thing}%
  },
  chronos period date/.style args={#1-#2-#3}{%
    chronos set date/.expanded={#1-#2-#3:otherthing}%
  },
  /chronos/.search also={/tikz},
  /chronos/.cd,
  start year/.store in=\chronosstartyear,
  start month/.store in=\chronosstartmonth,
  start day/.store in=\chronosstartday,
  end year/.store in=\chronosendyear,
  end month/.store in=\chronosendmonth,
  end day/.store in=\chronosendday,
  thing year/.store in=\chronosthingyear,
  thing month/.store in=\chronosthingmonth,
  thing day/.store in=\chronosthingday,
  otherthing year/.store in=\chronosotherthingyear,
  otherthing month/.store in=\chronosotherthingmonth,
  otherthing day/.store in=\chronosotherthingday,
  start date/.style args={#1-#2-#3}{%
    start year={#1},
    start month={#2},
    start day={#3},
  },
  end date/.style args={#1-#2-#3}{%
    end year={#1},
    end month={#2},
    end day={#3},
  },
  width/.store in=\chronoswidth,
  height/.store in=\chronosheight,
  period/.style={draw=gray},
  period event line/.style={draw=gray, -{Triangle[width=1.5pt, reversed, length=.75pt, fill=gray]}},
  period event/.style={anchor=north, fill=gray!25, draw=gray, rounded corners, align=center, font=\footnotesize},
  event line/.style={draw=gray, -{Triangle[width=1.5pt, reversed, length=.75pt, fill=gray]}},
  event/.style={anchor=north, fill=gray!25, draw=gray, rounded corners, align=center, font=\footnotesize},
  start date=1001-10-01,
  end date=1003-06-14,
  width=100mm,
  height=1pt,
  chronos date=1850-01-01,
  chronos period date=1851-01-01,
}
\NewDocumentCommand \chronosevent { O {} m O {} +m D () { -10pt-.5*\chronosheight } }
{%
  \scoped[on background layer]{\path [postaction={/chronos/event line, #1}, chronos date/.expanded={#2}] ({(\thechronosthingdate-\thechronosstartdate)*\chronosunit pt},0) -- +(0,#5) node [/chronos/event, #3] {\chronosthingday/\chronosthingmonth/\chronosthingyear\\#4};}
}
\NewDocumentCommand \chronosperiod { O {} m O {} m O {} +m D () { -10pt-.5*\chronosheight } }
{%
  \tikzset{%
    chronos date/.expanded={#2}, chronos period date/.expanded={#4}
  }
  \path [postaction={line width=\chronosheight, /chronos/period, #1}] ({(\thechronosthingdate-\thechronosstartdate)*\chronosunit pt},0) -- ({(\thechronosotherthingdate-\thechronosstartdate)*\chronosunit pt},0);
  \scoped[on background layer]{\path [postaction={/chronos/period event line, #3}] ({(.5*\thechronosotherthingdate+.5*\thechronosthingdate-\thechronosstartdate)*\chronosunit pt},0) -- +(0,#7) node [/chronos/period event, #5] {\chronosthingday/\chronosthingmonth/\chronosthingyear--\chronosotherthingday/\chronosotherthingmonth/\chronosotherthingyear\\#6};}
}

\makeatletter
\newcounter{subhyp}
\let\savedc@hyp\c@hyp

\newcommand{\normhyp}{%
  \let\c@hyp\savedc@hyp %
  \renewcommand\thehyp{\arabic{hyp}}%
}
\makeatother

\newcolumntype{L}[1]{>{\raggedright\let\newline\\\arraybackslash\hspace{0pt}}m{#1}}
\newcolumntype{C}[1]{>{\centering\let\newline\\\arraybackslash\hspace{0pt}}m{#1}}
\newcolumntype{R}[1]{>{\raggedleft\let\newline\\\arraybackslash\hspace{0pt}}m{#1}}

\input{numbers.tex}

\graphicspath{{figures/}}

\begin{document}

\title{Economies of scale and the equity-efficiency trade-off: Evidence from a piped water network in rural Bangladesh}\author{Md. Ahasan Habib and Anna Tompsett\thanks{\noindent \textbf{Correspondence: anna.tompsett@ne.su.se.} Habib: NGO Forum for Public Health, Dhaka, Bangladesh. Tompsett: Stockholm University and the Beijer Institute for Ecological Economics, Royal Swedish Academy of Sciences. We thank Guy Hutton and Dinesh Mehta for helpful discussions as well as conference participants at the WASH Economics Conference for their comments and suggestions. Md. Imran Ali, Md. Foysal Ahmed, Rubel Ahmed, Sakhawat Hossain, Arman Hossain, Mir Abu Raihan, and Tonmoy Kumer Saha provided outstanding research assistance. Funding: the International Growth Centre.}}

\maketitle

\normalsize

\vspace{-5ex} \begin{abstract}

\noindent Policymakers who set prices for public infrastructure often anticipate an equity-efficiency trade-off: while higher prices could increase revenues from fee-paying users, helping to offset costs, they may reduce take-up among the poor. But this trade-off is fundamentally altered in the presence of economies of scale, which are common in infrastructure networks. We elicit willingness to pay for a household-level piped drinking water connection in rural Bangladesh, where truthful revelation of preferences is incentivized by the opportunity to connect to a new pipeline network that we construct, and we use project cost data to measure economies of scale. Combining the two, we show that the cost-effectiveness-maximizing price---that which minimizes the cost per connected household---is zero. Economies of scale strongly dominate revenues raised by user fees. More generally, the two conditions needed for a zero price to maximize cost-effectiveness are important economies of scale and high sensitivity of demand to prices near zero, which is well documented across developing countries. Since a zero price naturally maximizes equity, we conclude that economies of scale can eliminate the equity-efficiency trade-off.

\end{abstract}

\newpage

\section{Introduction}\label{intro}

Setting prices for public infrastructure involves multiple, sometimes conflicting objectives. Policymakers want to achieve social goals, such as universal access, but they face many competing demands for limited resources. Setting prices higher increases revenues from each paying user but reduces take-up and prices out poorer households: an apparent equity-efficiency trade-off. An often neglected aspect of this problem, however, is that in the presence of economies of scale, take-up also affects the average cost of service provision. In this paper, we show empirically and theoretically that economies of scale can drive the cost-effectiveness-maximizing price to zero, eliminating the equity-efficiency trade-off.

The specific problem we study is how to price household connections to safe drinking water. We constructed a real piped water system in rural Bangladesh, which pumps safe groundwater from a deep borehole to elevated storage in a water tower, from where it flows by gravity to household taps through a pipeline network. We elicited willingness to pay for a piped water connection in \nWTPPart{} households, with truthful revelation of preferences incentivized by the opportunity to connect to the new pipeline. We then combine the data on demand with information about the pipeline network construction costs to derive the revenue-maximizing and cost-effectiveness-maximizing prices. Lastly, we derive theoretically the conditions under which a zero price can maximize cost-effectiveness and discuss when these conditions are likely to hold.

By focusing on cost-effectiveness, we depart from the classical literature on pricing for welfare maximization, which accounts for how different individuals value different benefits when trading off benefits against costs \parencite[e.g.,][]{H1938}. We argue that the focus on cost-effectiveness aligns closely with how contemporary policymakers make decisions. First, the policymaker cannot in general observe how individuals value benefits, typically considered equivalent to willingness to pay, making the welfare-maximizing policy an unrealistic benchmark. Second, in a highly unequal world, individual willingness to pay may reflect wealth more than differential valuation. For these reasons, policymakers often adopt strict equality in how they value benefits to different individuals or even explicitly pro-poor stances. For example, the US Office of Management and Budget recommends treating the value of a statistical life as uniform across individuals in regulatory cost-benefit analysis, while the funder GiveWell explicitly aims to identify and fund the interventions that are most cost-effective in saving or otherwise materially improving lives. Once we treat comparable benefits to different individuals as equivalent, maximizing cost-effectiveness becomes the natural way to choose between policy alternatives.

We elicited willingness to pay using a Becker--DeGroot--Marschak (\citeyear{BDM1964}) design. We asked households to report how much they were willing to pay to connect to the network. Households then received a randomly drawn price. If the price was equal to or below their reported willingness to pay, they would be able to connect at that price to the constructed system. We conducted the willingness-to-pay elicitation in two communities, one of which was later selected by lottery to receive the piped water network. We estimate a demand curve for a piped water connection that declines steeply with price, from \pctTakeupOneUSDPart\% of households at a price of US\$1 to \pctTakeupTwentyUSDPart\% at US\$20. The revenue-maximizing price is US\$\revMaxPriceUSDPart, but at this price only around a third of households connect. This is the apparent equity-efficiency trade-off.

We then turn to defining the cost function for the system using project data. The fixed costs total US\$\fixedCostUSD---the borehole, the pump, and the overhead storage system---while the per household connection costs---the pipeline and taps---are around US\$\varCostUSD{} per connected household. The fixed costs generate substantial economies of scale, implying that the average cost per connected household falls rapidly with every household that connects, up to the maximum capacity of about \capacityHH{} households.

We then combine our data on demand and costs to estimate the cost-effectiveness-maximizing price, or price that minimizes the cost per connected household. We find that this price is zero: any non-zero price causes demand to fall, reducing the number of households over whom the fixed costs are shared and increasing the average cost by more than the revenue raised.  In other words, a zero price would allow the policymaker with a fixed budget to connect the greatest number of households or equivalently to serve a fixed target number of households at lowest cost. The optimality of the zero price is robust across different samples and analysis choices. Holding the shape of the demand curve fixed and scaling up every household's willingness to pay by a constant factor, a positive price maximizes cost-effectiveness only when willingness to pay is more than \kScaleCEPositivePart{} times as high as reported in the elicitation.

The simple mechanism is that economies of scale in the piped water network are a substantially stronger driver of cost-effectiveness than revenues raised via user fees. At zero price, the cost per connected household is US\$\costPerHHFreeUSD. This compares to a theoretically ideal average net cost per connected household of US\$\costPerHHIdealUSDPart, achievable only if the policymaker could observe and charge each household their willingness to pay.

Lastly, our context is specific and our sample size is relatively modest, raising the natural question of to what extent our results are likely to generalize. We use a simple model to show the two conditions under which a zero price can maximize cost-effectiveness: significant economies of scale and high sensitivity of demand to prices near zero. The former varies across contexts and measurement approaches but appears common in rural networked infrastructure \parencite[e.g.,][]{LMW2020,CER2016}. The latter is a well-documented feature of demand in developing countries across a wide range of goods \parencite[e.g.,][]{D2014,BGG2012}.  An additional disadvantage of user fees is the cost of administering them \parencite{D2014b,CGH2026}, from which we abstract here.

Our results connect to several literatures. First, we contribute to the literature about how households value the benefits of infrastructure and in particular, access to safe drinking water. Simply asking households how much they are willing to pay may not elicit truthful responses. Households may understate their willingness to pay, if they believe that truthfully revealing a high willingness to pay will lead them to be charged higher prices, or overstate their willingness to pay, if they believe that provision of infrastructure depends on their reporting a high willingness to pay. This study joins a small set of papers that use experimental methods to elicit incentive-compatible valuations of safe water, specifically water treatments \parencite[e.g.][]{ABS2010,KMM2009,LMA2012,BFG2020}, delivery of safe water to households \parencite{BJS2026}, or community water sources \parencite{KLM2011}, where valuation is inferred using a travel cost model. Our study is unique in measuring incentivized willingness to pay for access to piped drinking water. Our findings nonetheless reiterate conclusions from those precedents: demand is very sensitive to price, and willingness to pay is often below the marginal cost of provision \parencite[see also][]{NHK2012}.

Second, we contribute to a literature describing infrastructure construction costs and in particular the presence of economies of scale. Whether studies find economies of scale in public service provision varies widely across contexts and types of public service, and depends on the methods researchers use \parencite{G-RL-PM-V2023}. We contribute a case study documenting economies of scale in rural networked infrastructure in a developing country. Notably, \textcite{LMW2020} experimentally recover a cost function for grid electricity rollout in rural Kenya and also find strong economies of scale, such that the average total cost per connection is minimized with universal free coverage. \textcite{CER2016}, using administrative and survey data, report similar conclusions for electricity networks in the Philippines. %

Third, we contribute to the literature on socially optimal ways to price public goods and services. Despite our focus on cost-effectiveness, our results align closely with the literature on welfare-maximizing price-setting. The undesirability of taxing goods with a high price elasticity of demand dates back to \textcite{R1927} and the case for subsidizing the fixed costs of infrastructure provision dates back to \textcite{H1938}. A closely related literature studies the problem of setting a uniform price scheme consisting of a connection fee and unit prices, with economies of scale in production---fixed and constant marginal costs---and a provider who must break even \parencite{AP1978,F1972}. This literature suggests that such a provider might prefer to increase unit prices above marginal costs---the canonical efficient solution---and reduce the connection fees: either to address equity concerns \parencite{F1972} or to avoid an inefficiently low share of households connecting, resulting in fewer users to share the fixed costs of provision \parencite{AP1978}. The idea that charging users can reduce cost-effectiveness when economies of scale are important is intuitive, but may often not be appreciated by policymakers \parencite[see, e.g.,][]{D2014b}. We provide empirical evidence of the importance of this mechanism and characterize the conditions under which a zero price can maximize cost-effectiveness.

The paper proceeds as follows: section \ref{context} describes the context, and section \ref{intervention}, the piped drinking water intervention. Section \ref{data} describes data collection and the design of the willingness-to-pay elicitation. Section \ref{WTP} describes willingness to pay for a pipeline connection and compares it to observed demand. Section \ref{cost_eff} combines estimates of demand with information about the cost of the intervention to evaluate the most cost-effective pricing strategy. Section \ref{model} develops a simple model that formalizes the conditions under which a zero price can maximize cost-effectiveness. Section \ref{conclusion} concludes.

\section{Context}\label{context}

The context for this research programme is the safe drinking water crisis in Bangladesh. Access to safe drinking water remains persistently low, despite years of effort by the Bangladeshi government and other organizations. The most recent systematically collected data \parencite{BBSUNICEF2019} suggest that 61\% of the Bangladeshi population---around 100 million people---lack access to a safely managed source of drinking water, as defined by the Sustainable Development Goals: free from contamination and available, on premises, when needed. In rural Bangladesh, the vast majority (92\%) use hand-pumped tubewells. Thirty million people use sources that are contaminated with arsenic, a naturally occurring pollutant in groundwater. Most people (136 million people or 82\%) have bacterial contamination in their drinking water, either from using a contaminated source or from recontamination during water transport or storage. %

Both exposure to faecal contamination in drinking water and arsenic contamination have potentially severe health effects. Exposure to faecal bacteria in drinking water increases the risk of diarrhoeal disease \parencite{LHH2015,KHC2022}, linked to long-term health and child development \parencite{P-UBC2014,PCA2017} and still the third-leading cause of child mortality worldwide \parencite{PKA2021}. %
Long-term exposure to arsenic also leads to a number of serious health impacts \parencite{HSC2014}.  Daily use of arsenic-contaminated water at the Bangladeshi standard of 50 parts per billion (ppb) is associated with an additional 1 in 100 lifetime risk of cancer \parencite{SLR2000}. Health impacts from exposure at lower levels are also documented, including levels below the more conservative WHO standard of 10 ppb \parencite{RNN2009}. The epidemic of arsenic-related disease in Bangladesh has been called ``the largest poisoning of a population in history'' \parencite{SLR2000}.

\begin{figure}[h]
\caption{Arsenic contamination in Bangladesh\label{fig:bangladesh_map_study_area}}
\begin{center}
\vspace{-2ex} \includegraphics[width=0.5\textwidth]{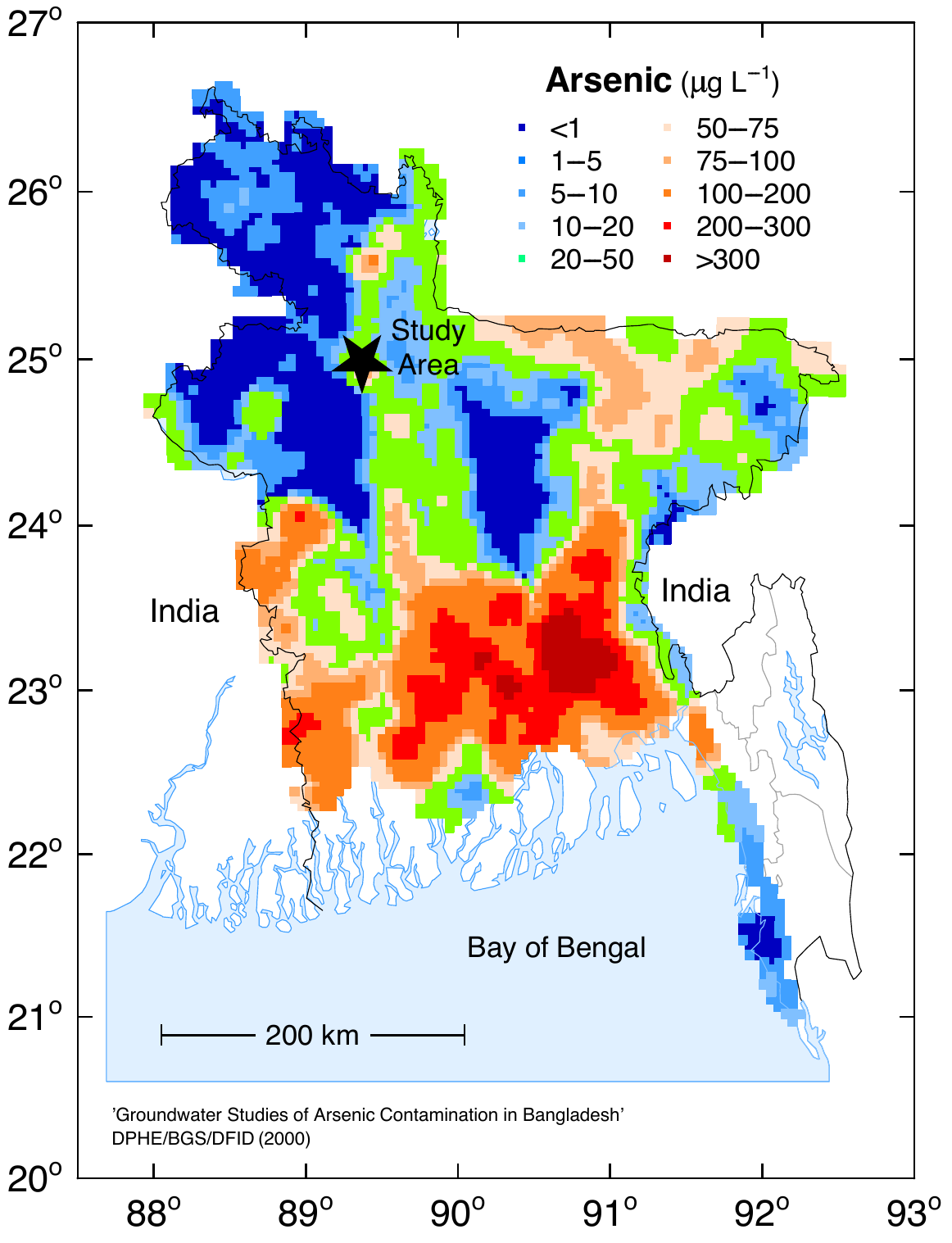}\\
\end{center}
{\vspace{-2ex} \footnotesize \emph{Notes} Arsenic contamination map reproduced with permission from \textcite{BGSDPHE2001}.  \par}
\end{figure}

The immediate context for our study is Shibganj Upazila in Bogra District in north-western Bangladesh (Figure \ref{fig:bangladesh_map_study_area}). Although the study area lies outside the epicentre of the arsenic contamination problem, government officials and national media reported high levels of arsenic in scattered pockets. The area had received few interventions to address the arsenic problem before a previous experiment to install arsenic-safe tubewells under different project implementation rules in some communities \parencite{CGH2021,CGH2026}.

In this study, we work in two communities---geographically contiguous groups of households with fewer than 250 households in 2015---that were randomly assigned to the control group in \textcite{CGH2021,CGH2026}. In these communities, we provided information about water quality in 2015---testing all water sources in the community for arsenic and faecal bacteria and providing the test results to water source caretakers---and had implemented two rounds of household surveys in a random sample of households, but we had not conducted any further interventions. The selection criteria for the two communities were: 1) that they were sufficiently large (more than 150 households) to ensure that building a pipeline would be reasonably cost-effective; and 2) that they had a geography that was amenable to the construction of a single pipeline system, i.e., with clusters that were not too geographically distant from one another and with a potential site for the water tower.

\section{Intervention}\label{intervention}

The physical intervention we study is a piped water network that provides households with individual water connections (taps). We constructed the piped water network in one of the two study communities, selecting which would receive the piped water network in a public lottery, conducted after the willingness-to-pay elicitation.

\begin{figure}[h]
\caption{Map of pipeline network\label{fig:connection_map}}
\begin{center}
\vspace{-2ex} \includegraphics[height=0.8\textheight]{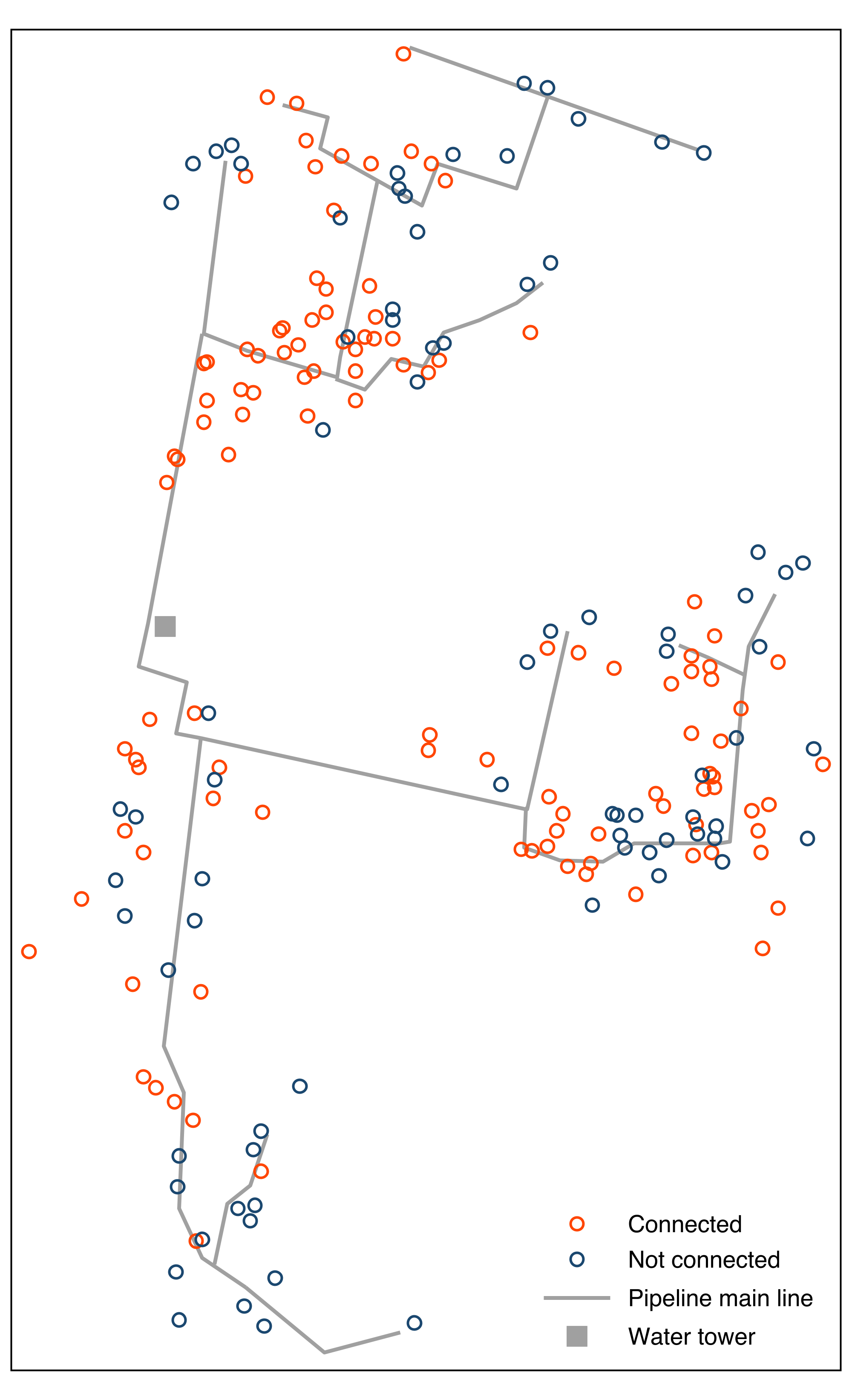}\\
\end{center}
{\vspace{-2ex} \footnotesize \emph{Notes} $N$ = \nMapHH{} households that consented to participate in household surveys in the community that received the pipeline. Household location data based on GPS coordinates collected in 2023.  Background imagery and geographical identifiers are not shown to preserve the anonymity of the study community.  \par}
\end{figure}

The piped water network has three key features: 1) a borehole to a safe aquifer---an underground layer through which groundwater flows---that is sufficiently deep to be isolated from sources of both bacterial and arsenic contamination; 2) a motorized electric pump to an overhead tank (Appendix Figure \ref{fig:water_tower}); and 3) a simple gravity-fed pipeline distribution system.  A sensor in the overhead tank detects when water levels are low and automatically initiates pumping, so that water can flow continuously through the pipeline network. In addition to the household connections, the pipeline network has three community taps, by the mosque in each cluster. This design ensures that any household living in the community can access drinking water from the system. Figure \ref{fig:connection_map} maps the constructed pipeline.

The network has a theoretical maximum capacity of \capacityHH{} households, each of whom can be served with their own private tap.  At the time of construction, our project documents record that the system served \nConnectedTotal{} households, with pipelines laid to connect a total of \nPipelineLaid{} households. The total installation cost was US\$\totalCostUSD{} at prevailing exchange rates, or US\$\costPerLaidUSD{} per household with pipeline laid.\footnote{Throughout, cost data and prices are converted from Bangladeshi taka to US\$ using the prevalent exchange rate of 119.5 BDT:1US\$.}  This breaks down to a cost of US\$\fixedCostUSD{} for the borehole, pump, and overhead storage system (the fixed costs) and a per household cost of the pipe network of US\$\varCostUSD{} (the variable costs). Managing the piped water system requires monthly costs of around US\$\runCostMonthUSD{}, or US\$\runCostMonthPerHHUSD{} per currently connected user household, to be funded by water fees from tap owners, but these are small compared to the system's installation costs.

\begin{figure}[th]
\caption{Costs of pipeline\label{fig:costs}}
\begin{center}
\vspace{-2ex} \includegraphics[width=0.7\textwidth]{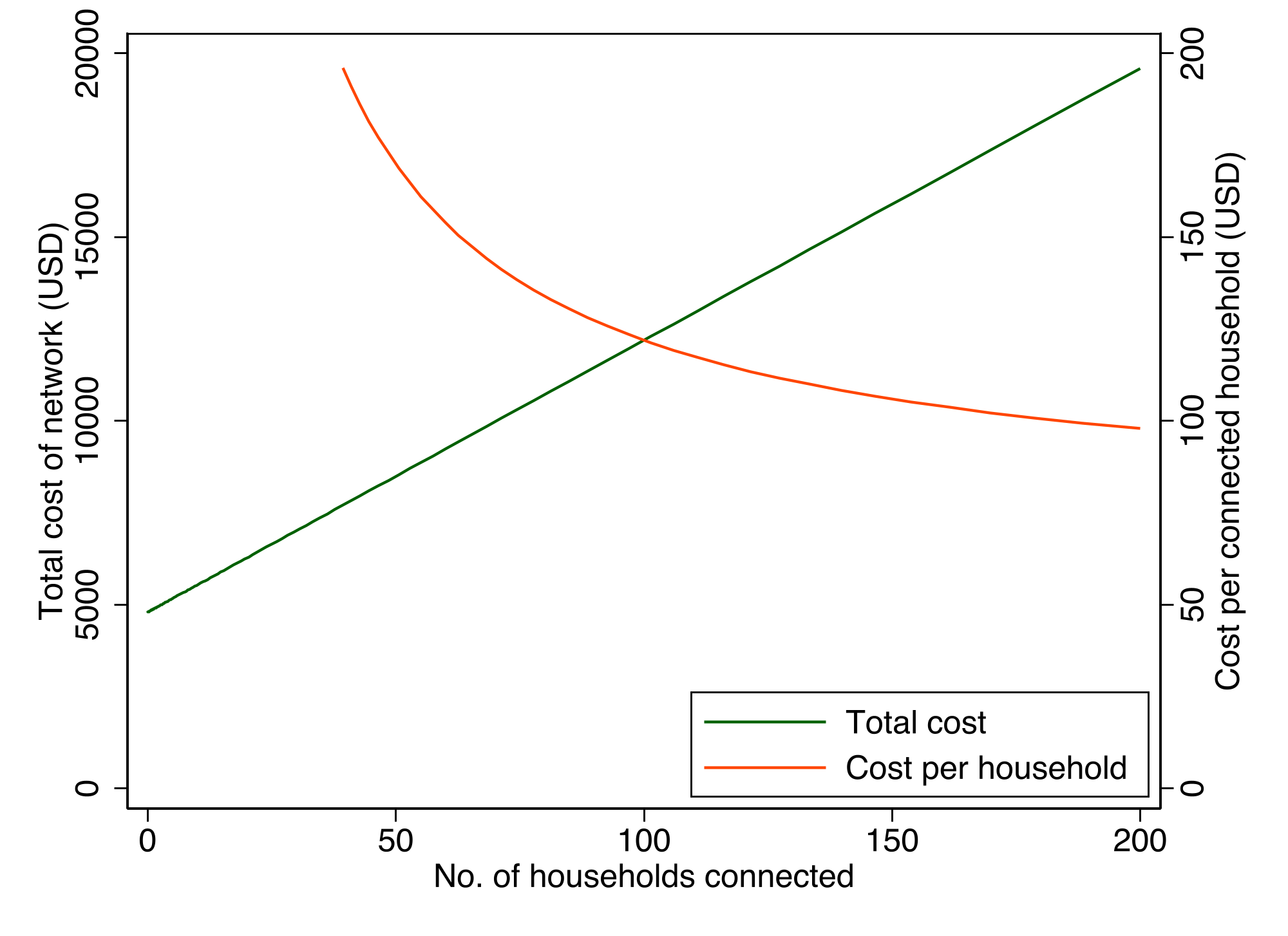}\\
\end{center}
{\vspace{-2ex} \footnotesize \emph{Notes} Graph shows total costs and costs per household connected of a local pipeline network, up to a maximum capacity of \capacityHH{} households. The graph of cost per household is truncated at US\$200 for readability. \par}
\end{figure}

The fixed costs create economies of scale whereby the cost per household falls with increasing numbers of connected households (Figure \ref{fig:costs}). We assume that the total variable costs scale linearly with the number of households connected. In practice, this assumption probably underestimates economies of scale. When a larger share of households in a community connect, the distance between the connected households reduces, with each connected household requiring a smaller marginal increment of pipeline and excavation work, meaning that marginal costs also decline.

The pipeline is designed to deliver safe drinking water into people's homes, reducing the risk of recontamination by avoiding the need for transport and storage. Pipelines also offer advantages beyond improvements in water quality.  A large share of households (\pctInterested\%)  in our study communities express interest in a household piped water connection. These households cited, on average, \meanAdvantages{} potential benefits of a piped water connection.  Among the advantages cited, households emphasised the labour-saving nature of a household connection. Notably, a substantial share of households cited the reduced need for women to move outside the home to collect water, a striking feature of the data given the emphasis in the policy literature on the idea that better access to water empowers women (Appendix Figure \ref{fig:advantages}).

\clearpage

\section{Data collection}\label{data}

This section describes the organization of the project. We use data collected in 2015, 2023, and 2024. The pipeline network was constructed after data collection in 2023 and completed immediately before data collection in 2024.

\paragraph{Household data} We conducted a household census in 2023, during which we updated administrative household lists that we first assembled in 2015.\footnote{Of \nListedHH{} households listed on the administrative lists from 2015, we confirmed that \nListedFound{} were present in the study communities. Of those not living in the study communities: \nListedMigrated{} had migrated; in \nListedNoSurvivors{}, all household members had passed away; and \nListedMislisted{} were mis-listed. We also registered \nCensusNew{} new households and updated information about the current household head for \nHeadChanged{} households.}  We then attempted to interview all households in both study communities within a spatial area we had delineated as being feasible to connect with a pipeline. A total of \nSurvey{} households (\pctSurveyEligible\% of eligible households) agreed to participate in the household survey.\footnote{Among the \nCensusHH{} households we had identified during the household census, we established that a further \nCensusMigrated{} had migrated and \nCensusMerged{} households had merged with another household, for example, an elderly relative had moved in with a younger relative.  Out of the remaining \nCensusFound{} households, we excluded a further \nOutsideArea{} who lived outside the anticipated feasible pipeline area. We sought consent to interview the remaining \nEligible{} households, among whom \nDeclinedAll{} households declined to participate.} We exclude households who declined to participate in interviews from all subsequent analyses.  The household survey in 2023 included modules on household information, all water sources used by the household, knowledge about safe drinking water, household assets, and the WTP elicitation, described in detail below. We resurveyed households in 2024, after construction of the pipeline. Among the \nSurvey{} households surveyed in 2023, we successfully resurveyed \nResurvey{} households in 2024 (\pctResurvey\%).\footnote{In \nResurveyNoSurvivors{} households, no household members survived; \nResurveyDeclined{} household declined to participate; and \nResurveyNotFound{} others could not be surveyed, primarily because no one was at home during the survey window.}

\paragraph{Water source data} We compiled data on all water sources in each community in 2015 and 2023. In 2015, we identified \nWellsListedCensus{} water sources. Between 2015 and 2023, \nWellsCensusGone{} of these sources ceased to exist and \nWellsCensusNew{} new sources were identified. Almost all sources recorded in or before 2023 were privately owned shallow tubewells. In the 2023 and 2024 household surveys, we recorded every source used by each surveyed household for drinking or cooking. In 2023, \nSourcesUsedBase{} sources were used by at least one surveyed household. In 2024,  \nSourcesUnusedEnd{} of the sources used in 2023 were no longer used by any surveyed household;  \nConnected{} surveyed households in the treated community reported having a piped water connection;\footnote{Three households who our project documents show as receiving a connection did not report using a piped connection in the survey. For all other surveyed households, self-reported pipeline use aligns with project records.} and \nHHNewTubewell{} surveyed households reported using a new tubewell, \nNewTubewellsTreated{} in the community receiving the pipeline and  \nNewTubewellsControl{} in the control community.

Each water source is recorded in our data with a text description and a photograph, allowing us to match individual households to the water sources they use. We tested water quality in all sources in 2015, and all sources used for drinking and cooking in 2023. We offered households the opportunity to participate in the water quality testing programme separately from the household survey. In practice, no household who declined to participate in the survey consented to the water quality testing programme. One household who consented to the survey declined to participate in the water quality testing programme. In this paper, we use data on arsenic contamination, measured using arsenic field tests. Arsenic field tests are less accurate than laboratory testing \parencite{RRW2020} but they are quick and inexpensive, and they obviate the need to send samples to external laboratories where we have only limited control over tracking procedures. In total, we collected baseline water quality data for 353/355 water sources used by any household in 2023.\footnote{One observation is missing from the household who did not consent to water quality testing; one household's record could not be matched to its water source test data.}

\paragraph{Demand elicitation design}

To measure demand for an individual private connection, we use a Becker--DeGroot--Marschak (\citeyear{BDM1964}) or BDM design to elicit truthful revelation of the maximum price households are willing to pay for a piped water connection.

We designed the BDM elicitation to be incentivized through a household-level price lottery. Respondents answer a series of eight yes-or-no questions that iteratively narrow in on the maximum amount a household is willing to pay for a connection. The dynamic price list ensures that we can pin down willingness to pay (WTP) to a narrow interval (of width US\$\wtpIntervalUSD) while keeping the question format simple and straightforward (Appendix Table \ref{tab:dynamic_price_list}). After confirming the WTP, respondents select one of five envelopes containing different price vouchers. The vouchers can list prices between zero and a maximum of just above US\$\voucherMaxUSD, the expected cost per connected household. Vouchers were distributed approximately uniformly across \nVoucherPrices{} possible prices, at intervals of about US\$\voucherStepUSD, but with more vouchers at zero and fewer at the highest prices (Appendix Figure \ref{fig:voucher_design}). We did not disclose the distribution of voucher prices in advance. If the price voucher was equal to or below the stated WTP, the household kept the voucher. The voucher could then be redeemed if a piped water network was constructed in the community for a household piped water connection at the listed price. We explained to households that they would only be able to redeem their vouchers if we constructed a pipeline in their community and that their community had a 50\% chance of being selected to receive the pipeline.  The vouchers were personalized and non-tradable, and could only be redeemed for a piped water connection on the named household's land.

To ensure that households understood the WTP elicitation, we first conducted a separate WTP elicitation for a familiar good: a bar of soap.  We also coded dynamic questions to first confirm the WTP and then to confirm that respondents understood what would happen if they drew a price just above and just below their stated WTP. Among the \nSoapBelowWTP{} households who drew a price below their WTP for the soap, \pctSoapBought\% purchased the soap at the voucher price, consistent with good understanding of the mechanism.

Because the price vouchers are drawn on the fly, one might be concerned about whether the random assignment of prices to households occurred as intended. We show that the distribution of vouchers drawn is largely consistent with the expected distribution, with some excess or missing mass at a number of prices likely due to imperfect shuffling of the physical vouchers (Appendix Figure \ref{fig:voucher_design}). We also confirm that the prices drawn do not correlate with household wealth for either the soap elicitation or the piped water elicitation (Appendix Figures \ref{fig:price_wealth_soap} and \ref{fig:price_wealth}), and that the prices drawn for soap are orthogonal to the prices drawn for the piped water connection (Appendix Figure \ref{fig:price_soap_pipe}).

We sought separate consent for the WTP elicitation, so that households who did not wish to participate in the survey or water quality testing could still choose to participate in the WTP elicitation.  However, no household who declined to participate in the household survey chose to participate in the WTP elicitation. Among households who consented to participate in the household survey, \nWTPDeclined{} declined to participate in the WTP elicitation.

Although the BDM approach is widely implemented, it was unclear ex ante how households would experience the WTP elicitation. To shed light on this, we asked households either immediately before or after the elicitation whether any household members were currently experiencing stress or emotional distress. We randomly assigned households to either be asked the question before or after the elicitation.  Households asked the question immediately after the elicitation were less likely to report distress than households asked the same question immediately before the elicitation (Appendix Figure \ref{fig:WTP_distress}). We interpret this as suggesting that households broadly experienced the WTP elicitation in a positive way.

Willingness to pay for a good should partly reflect household characteristics that are common to demand for all consumer goods, such as wealth. As such we normally expect WTP for different goods to be positively correlated. WTP for a piped water connection is indeed positively correlated with WTP for the bar of soap (Appendix Figure \ref{fig:pipe_vs_soap}).

\section{Willingness to pay}\label{WTP}

\begin{figure}[h]
\caption{Demand and average revenue\label{fig:rev_max_usd}}
\begin{center}
\vspace{-2ex} \includegraphics[width=0.8\textwidth]{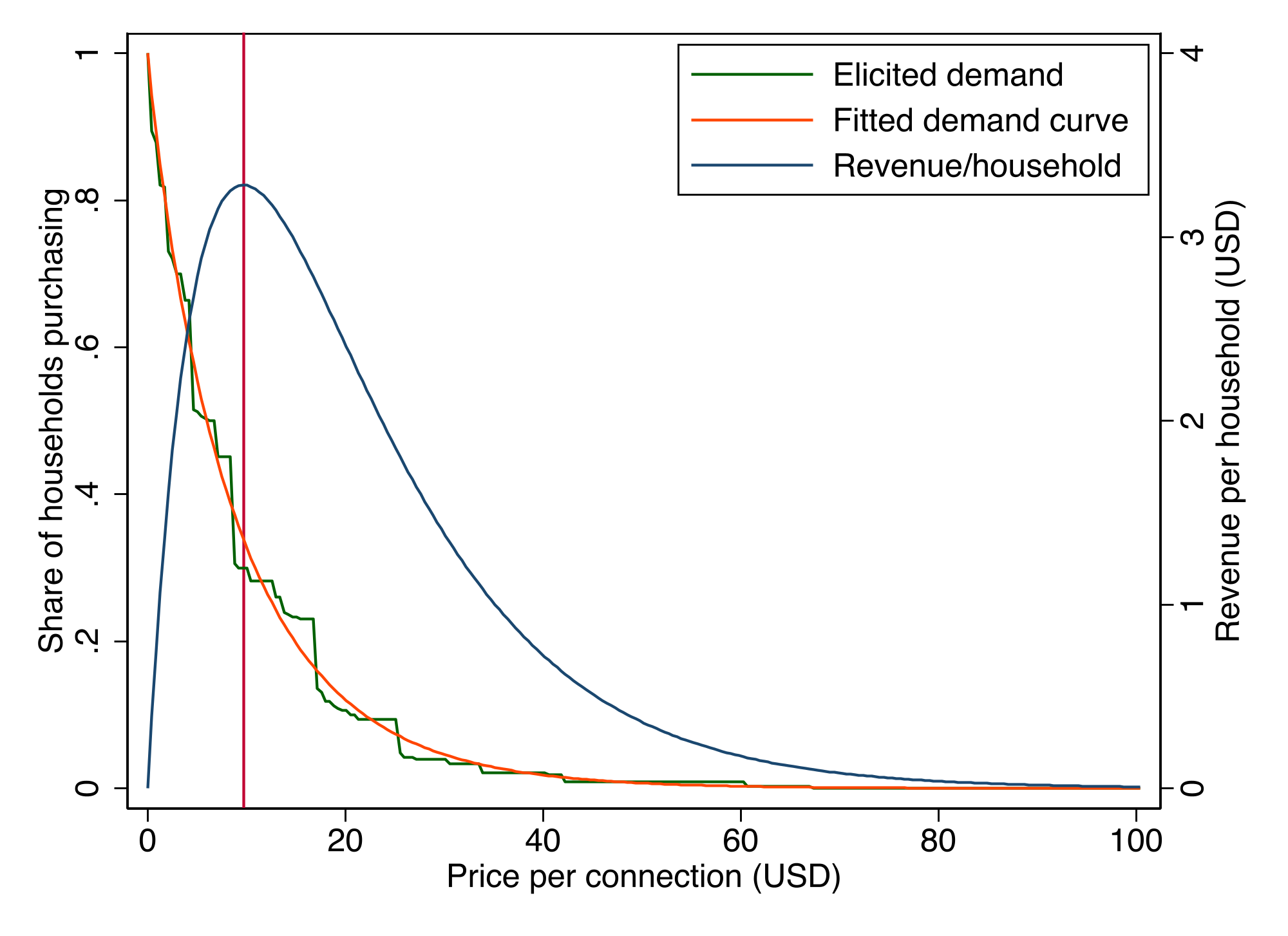}\\
\end{center}
{\vspace{-2ex} \footnotesize \emph{Notes} $N$ = \nWTPPart{} eligible households that participated in the willingness-to-pay elicitation. Raw data from willingness-to-pay elicitation shown in green.  Weibull function fitted to elicited demand data shown in orange ($e^{-\weibullAPartUSD \, p^{\weibullBPartRound}}$, where $p$ is the price in US dollars).  Revenue per household, shown in blue, is the price multiplied by the share of households purchasing at that price under the fitted Weibull function. The red vertical line marks the revenue-maximizing price (US\$\revMaxPriceUSDPart). \par}
\end{figure}

\paragraph{Estimated demand} Figure \ref{fig:rev_max_usd} shows the results of the willingness-to-pay (WTP) elicitation for the households that agreed to participate.  The raw data are shown in green.  Around \pctWTPZeroPart\% of participating households declared a WTP of zero. The median household has a WTP of around US\$\wtpMedianUSDPart.  The maximum WTP declared by any household is US\$\wtpMaxUSDPart, but this is the extreme of a skewed distribution: the 90th percentile is around US\$\wtpNinetyUSDPart.  The demand curve is quite well approximated by a Weibull function (shown in orange).  Table \ref{tab:summary_demand} summarizes the properties of the demand curve.

A policymaker wishing to maximize revenue, faced with this demand curve, would set a price of approximately US\$\revMaxPriceUSDPart. How average revenue across all households varies as a function of price is shown in Figure \ref{fig:rev_max_usd} in blue. For low prices, relatively many households connect, but each connecting household contributes little in revenue. For high prices, each household contributes a greater amount, but few households connect. At the revenue-maximizing optimum, however, only around a third of households would connect, a classic equity-efficiency trade-off (Table \ref{tab:summary_demand}, columns 3 and 4).

Around \pctWTPDeclined\% of surveyed households (\nWTPDeclined{} of \nSurvey) declined to participate in the WTP elicitation. One interpretation of this choice is that these households had a WTP of zero. If we assume this is the case, the overall estimated demand curve falls (Table \ref{tab:summary_demand}, row 2; see also Appendix Figure \ref{fig:rev_max_usd_all}). However, the estimated revenue-maximizing price barely shifts and the equity-efficiency trade-off, if anything, intensifies.

Demand varied slightly across the two communities. At the time of the WTP elicitation, it was not yet determined which community would receive the pipeline. In the community that eventually received the pipeline, demand was somewhat higher among households that participated in the WTP elicitation than in the same group in the eventual control community (Table \ref{tab:summary_demand}, rows 3 and 5; Appendix Figure \ref{fig:demand_by_village}; Kolmogorov--Smirnov test $p$ = \ksPVillagePart). In the eventually treated community, a larger share of households also declined to participate in the elicitation (\pctWTPDeclinedTreat\% compared to \pctWTPDeclinedCtrl\%). If we assign all households that declined to participate a WTP of zero, the differences between the demand curves attenuate (Table \ref{tab:summary_demand}, rows 4 and 6; Kolmogorov--Smirnov test $p$ = \ksPVillageAll; see also Appendix Figure \ref{fig:demand_by_village_all}). In both communities, the estimated demand curve yields a similar revenue-maximizing price, regardless of how we treat the households who chose not to participate in the WTP elicitation.

\paragraph{Observed demand} After the WTP elicitation and assignment to treatment, we eventually capped the required payment at US\$\capPriceUSD{} (1000 BDT). Households with a voucher below this value could purchase at the voucher price; all other households could purchase a connection for 1000 BDT. We made this change to the design because demand revealed in the WTP elicitation fell faster than we expected when we designed the elicitation (Appendix Figure \ref{fig:obs_vs_exp}). In itself, it is not surprising that demand differed from our expectations: the lack of information about the demand curve motivated our experiment, so our expectations were purely speculative. But a consequence of the lower-than-expected demand was that only \nVoucherRetained{} households retained a voucher after the WTP elicitation.  Building a pipeline to serve only \nVoucherRetained{}  households would have been very inefficient and would have undermined other research goals. Households also expressed concern about potential unfairness if households paid very different amounts for the same good.

\clearpage

\begin{landscape}
\begin{table}[p]
\caption{Demand for a household connection: summary\label{tab:summary_demand}}
\begin{center}
\small
\input{tables/summary_demand.tex}
\end{center}
{\footnotesize \emph{Notes} Sample excludes the \nDeclinedAll{} eligible households that declined to participate in any component of the study. Panel A: data from the willingness-to-pay (WTP) elicitation. Rows labelled ``participants'' include only households that participated in the WTP elicitation; rows labelled ``WTP = 0 if declined'' also include households that participated in the household survey but declined to participate in the WTP elicitation, assigning them a WTP of zero. Column 2 shows the share of households with a WTP of at least 1000 BDT. Columns 3--5 are based on a Weibull demand curve fitted separately to each sample. The cost-effectiveness-maximizing price minimizes the net cost per connected household over a grid of possible prices. Panel B: observed take-up at a price of 1000 BDT in the treated community, excluding households that drew a free connection in the price lottery. In the ``all eligible'' row, the remaining participants are reweighted by the inverse of the probability of drawing a non-zero price, so that participants and households that declined the WTP elicitation are represented in their sample proportions. \par}
\end{table}
\end{landscape}

In principle, since the price cap was only affected by individual responses to the extent that they affected the estimated aggregate demand function, this change does not invalidate the incentivized design. Even if an individual anticipated that the community's aggregate response could influence the final price we set, their individually optimal response would still be to reveal something very close to the truth. This is because the benefits of understating their willingness to pay accrue to everyone; the benefits of telling the truth only to themselves. The temptation to free ride helps incentivize truthful revelation.

In total, our project documents record that \nConnectedTotal{} households connected to the pipeline network at the time of construction, just before we collected data in 2024. Of these, \nTreatedSampleConnected{} are among the \nTreatedSample{} households in our study sample in the treated community; the remainder are households outside the study sample, including new households formed from the division of other households after our household listing.  Most connecting households paid 1000 BDT (US\$\capPriceUSD), the maximum price. Twenty households who drew a voucher for a free connection during the WTP elicitation received a connection for free (100\% take-up).  Of the \nVoucherEightHundredRetained{} households who retained a voucher for a connection at 800 BDT, the \nVoucherEightHundredRetainedConn{} who connected in practice also paid 1000 BDT.

An aggregate take-up rate of \pctObsTakeupThousandPart\% at 1000 BDT aligns very well with the predicted average take-up rate of \pctPredTakeupThousandPart\% at 1000 BDT in the sample of households who participated in the willingness-to-pay elicitation (Table \ref{tab:summary_demand}).\footnote{The calculation of observed take-up at 1000 BDT naturally excludes households that drew a zero price in the lottery.  Since the zero price was randomly assigned among households who participated in the WTP elicitation and price lottery, excluded households are not systematically different from other households who participated in the price lottery.} Among the households who declined to participate in the WTP elicitation, take-up was \pctTakeupNonPart\%, only slightly lower than for those that did participate, resulting in an overall take-up rate of \pctObsTakeupThousand\%.\footnote{We reweight the households that participated in the WTP elicitation by the inverse of the probability of drawing a non-zero voucher when we calculate observed demand in the full sample to ensure representativeness.}

Unexpectedly, however, predicted take-up based on responses to the WTP elicitation has very little predictive power over final take-up.\footnote{One could be concerned that this results from a mismatch of the observed take-up data to the household survey data, but we confirm that we correctly match the \nFreeVoucherTreated{} households who drew a price of zero in the price lottery to their take-up data, suggesting that this cannot explain the lack of correspondence, although we cannot rule out the possibility that matching was more rigorous for these households.} Take-up at a price of 1000 BDT is \pctTakeupLowWTPPart\% in households with a WTP of less than 1000 BDT and \pctTakeupHighWTPPart\% in households with a WTP of 1000 BDT or higher, a small and statistically insignificant difference ($p$ = \pTakeupWTPPart).  Including the households who chose not to participate in the WTP elicitation, assigning them a zero WTP, does not increase predictive power (Appendix Table \ref{tab:takeup_prediction}).

Examining the distribution of take-up suggests a possible explanation. Take-up is spatially clustered (Figure \ref{fig:connection_map}), and the clusters of take-up coincide with clusters of relatively high arsenic contamination (Appendix Figure \ref{fig:arsenic_map}). Take-up is higher for households with higher arsenic contamination: take-up is \pctTakeupNoArsenic\% among households with no detectable arsenic contamination at baseline, \pctTakeupArsenicTen\% among households with arsenic contamination above the WHO threshold (10~ppb) but below the Bangladeshi threshold (50~ppb); and \pctTakeupArsenicFifty\% among households with arsenic contamination above 50 ppb. We can strongly reject the null hypothesis of equal take-up across these groups ($p$ < 0.01, see Appendix Table \ref{tab:arsenic_takeup}). The most common reason given by those who did not take up a pipeline connection was that they were satisfied with their own source (Appendix Figure \ref{fig:reasons_no_connect}).

In contrast, arsenic contamination is \emph{not} correlated with take-up predicted by responses in the WTP elicitation; if anything, the pattern is reversed, with lower predicted take-up for higher levels of arsenic contamination (Appendix Table \ref{tab:arsenic_takeup}).  Because information about arsenic contamination was not provided to households until after the WTP elicitation, this suggests that a reasonable explanation for the differences between predicted and observed take-up is that the information about arsenic we provided at the end of the baseline survey resolved some uncertainty, meaningfully changing individual households' demand. The reason that observed take-up is still similar to predicted take-up based on the WTP elicitation is because those who found out their sources were uncontaminated took up connections at lower rates than predicted while those who found out their sources were contaminated took up connections at higher rates.

The strong response to information is somewhat surprising because we had already provided communities with extensive information about arsenic contamination, albeit up to eight years previously. Arsenic contamination is relatively stable, so test results from 2015 should still have been informative in 2023 (Appendix Figure \ref{fig:arsenic_persistence}). Yet the information we provided on arsenic contamination after the WTP elicitation largely seems to have been unknown to the households at baseline. About 70\% of households reported that their sources had been tested, but \pctSafeIfTested{}\% reported that their source was safe, if tested. Only \pctPaintedGreen{}\% of sources were painted green to indicate safety. Household responses were only very weakly correlated with whether the source later tested positive for arsenic (Appendix Table \ref{tab:beliefs_arsenic}).\footnote{Measured arsenic contamination in 2023 is uncorrelated with whether or not a household reports a source to have been tested or whether a source has been reported to have been tested and found safe. Measured arsenic contamination is somewhat correlated with whether a household reports a source to be safe among the sources that are reported to have been tested and whether or not a source is painted green, but the differences are small in magnitude.} Households had not abandoned sources with higher contamination in 2015 at higher rates, and although they were more likely to report them having been tested previously, they were only weakly less likely to report them being safe if tested (Appendix Table \ref{tab:beliefs_arsenic_2015}). Overall, these results suggest that households did not retain the information provided in 2015 about arsenic, possibly because the information was not especially useful in the absence of opportunities to act on it. In contrast, the information provided in 2023 was immediately followed by an opportunity to connect to an arsenic-free water source, making it both salient and actionable. Another explanation that is also consistent with the results we observe is that our field staff exerted more effort in convincing households with higher arsenic contamination to connect to the pipeline.

\section{Cost-effectiveness}\label{cost_eff}

Combining these data with information on the costs of the pipeline network, we can determine the optimal strategy for a policymaker who wishes to maximize cost-effectiveness in their intervention, that is, to serve the greatest number of households at the lowest possible cost. We model demand for connections as a function of price using data from the WTP elicitation, as shown in Figure \ref{fig:rev_max_usd}, and calculate the total cost of the system and the cost per connected household, before and after revenues, at fine intervals for prices on a grid between zero and an upper limit at which take-up is close to zero. Strikingly, the most cost-effective approach is to offer connections for free. This is because economies of scale strongly dominate the revenue-raising effect of charging for household connections.  In fact, given the demand curve based on the WTP elicitation, revenues raised by user fees are almost irrelevant in determining costs per household.

\begin{figure}[h]
\caption{Costs as a function of connection price\label{fig:cost_w_price}}
\begin{center}
\vspace{-2ex} \includegraphics[width=0.8\textwidth]{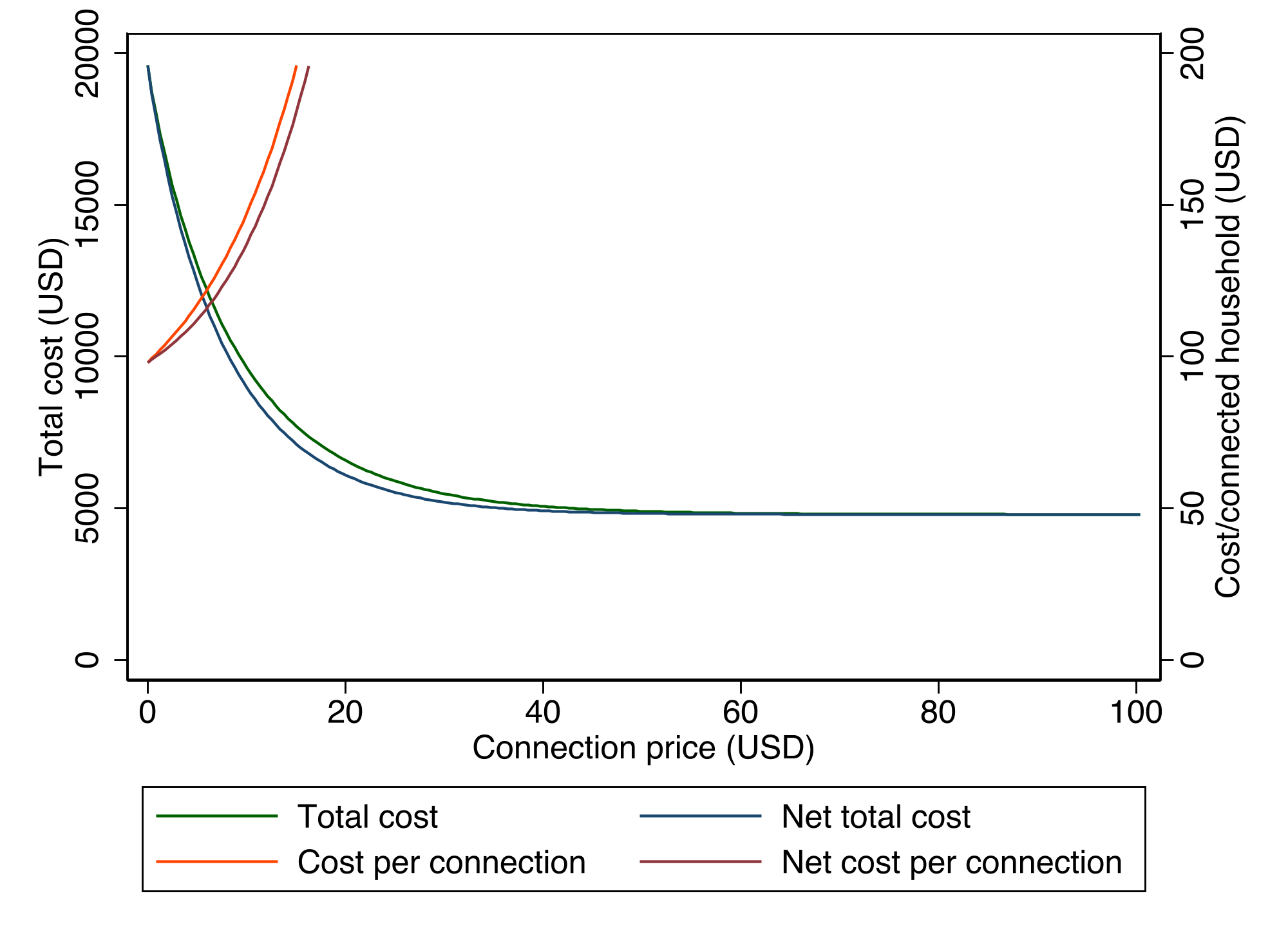}\\
\end{center}
{\vspace{-2ex} \footnotesize \emph{Notes} Graph shows total cost of the piped water network as a function of the connection price (green) and total cost net of revenues raised by households paying to connect to the network (blue).   Costs per household shown in orange; costs per household net of revenue shown in red. Costs per household are truncated at US\$200 for readability.  Demand is the Weibull curve fitted to households that participated in the WTP elicitation (Figure \ref{fig:rev_max_usd}). \par}
\end{figure}

A uniform price is a coarse instrument, but a hypothetical policymaker who could perfectly differentiate user fees based on willingness to pay would not be able to implement the project at much lower cost than the zero price strategy.  Implementing the programme with free connections would cost US\$\costPerHHFreeUSD{} per connected household. A policymaker who could charge individual households their willingness to pay would raise average revenues of US\$\wtpMeanUSDPart{} per household, resulting in a programme cost per connected household of US\$\costPerHHIdealUSDPart, only a \pctIdealSavingPart\% saving relative to free provision. This difference would naturally narrow if we accounted for the costs of administering the user fees.

\begin{figure}[h]
\caption{Revenue-maximizing and cost-effectiveness-maximizing prices with higher willingness to pay\label{fig:wtp_scale}}
\begin{center}
\vspace{-2ex} \includegraphics[width=\textwidth]{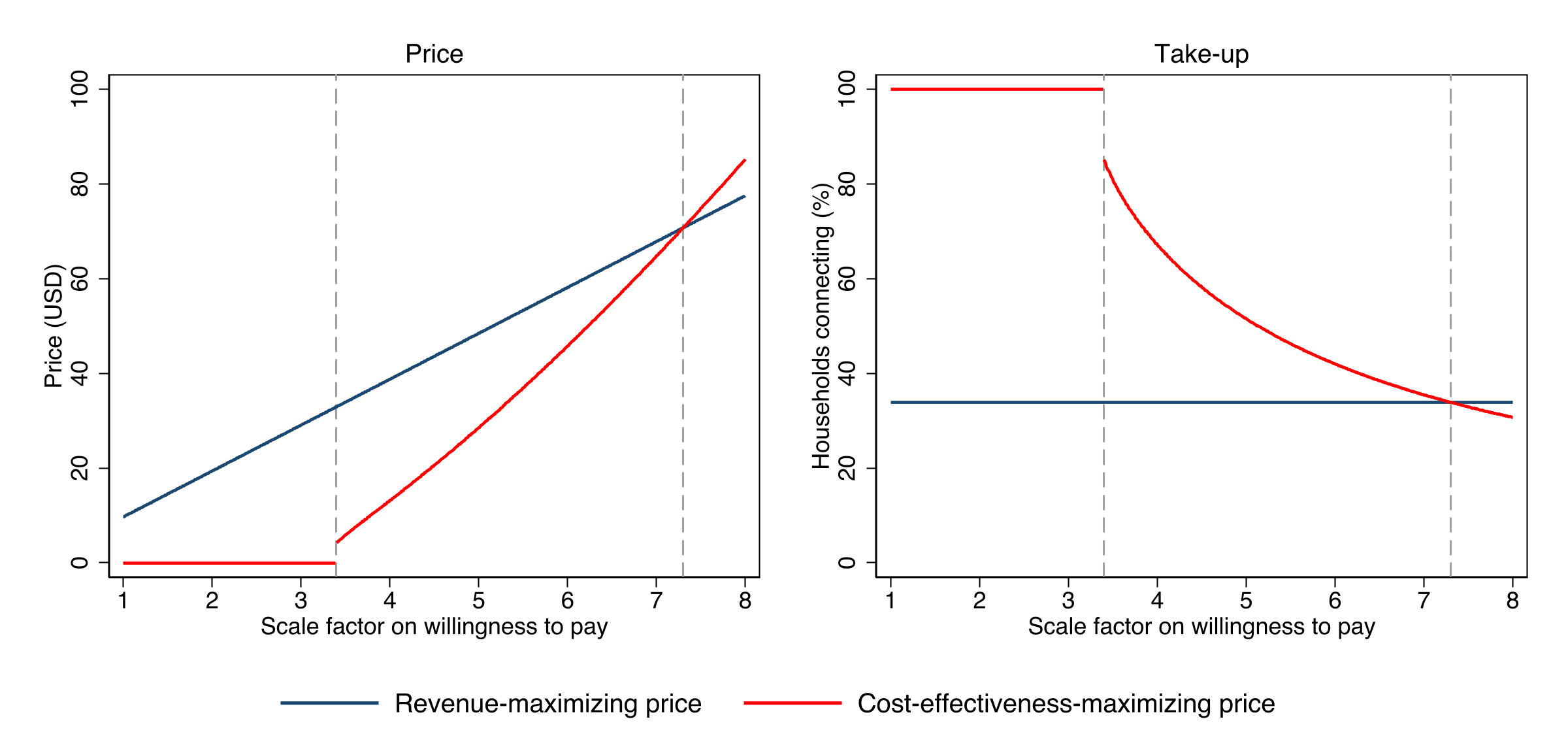}\\
\end{center}
{\vspace{-2ex} \footnotesize \emph{Notes} Left: revenue-maximizing price (blue) and cost-effectiveness-maximizing price (red) when the willingness to pay of every household that participated in the WTP elicitation is multiplied by a scale factor. Right: share of households connecting at each price.  Dashed lines mark the scale factors at which the cost-effectiveness-maximizing price first exceeds zero (\kScaleCEPositivePart) and at which it equals the revenue-maximizing price (\kScaleCEEqualsRevPart). \par}
\end{figure}

A reasonable question is how much demand would have to change before the zero price no longer maximizes cost-effectiveness. We generate hypothetical demand curves where willingness to pay is multiplied by factors up to 8. Each of these generates a new Weibull demand curve with the same shape parameter as the elicited demand curve (Appendix Figure \ref{fig:wtp_scale_demand}).  We calculate the revenue-maximizing and cost-effectiveness-maximizing prices for each of these scenarios (Figure \ref{fig:wtp_scale}). Because the shape of the demand curve is held constant, the revenue-maximizing price simply scales with the multiplication factor, and take-up at the revenue-maximizing price is always around one third. The cost-effectiveness-maximizing price is zero for scaling factors below \kScaleCEPositivePart{} and then jumps to a positive price. The  cost-effectiveness-maximizing price is equal to the the revenue-maximizing price at a scaling factor of \kScaleCEEqualsRevPart.  At a scaling factor of \kScaleCEPositivePart{}, more than 1 in 10 households are modelled to connect at a price that only 1 in \nWTPPart{} households stated as their willingness to pay in the elicitation (US\$\wtpMaxUSDPart).

\section{A simple model for pricing infrastructure connections}\label{model}

Consider a community-level infrastructure system to serve a community with $\overline{M}$ households. Let the total cost of the infrastructure system be $F+cM$, where $F>0$ is the fixed cost of serving the community, $c$ is the marginal cost of connecting a household, and $M \leq \overline{M}$ is the number of households connected. Let demand for connections $M\left(p\right)$ vary as a function of price $p$. For simplicity, assume that take-up is universal when the price is zero ($M\left(0\right) = \overline{M}$), that there exists a finite price $\overline{p}$ at which demand for connections falls to zero ($M\left(\overline{p}\right) = 0$), and that $M\left(p\right)$ is continuous and $M'\left(p\right) < 0$ for $p\leq\overline{p}$.

A policymaker wishes to maximize efficiency in spending on infrastructure, equivalent to minimizing the net public expenditure per connected household, specifically: \begin{align*}
\min_p A\left(p\right)=\frac{F+cM\left(p\right)-pM\left(p\right)}{M\left(p\right)}=\frac{F}{M\left(p\right)}+c-p, 0 \leq p < \overline{p}
\end{align*}

The associated Lagrangian is: \begin{align*}
\mathcal{L}(p,\lambda)=\frac{F}{M(p)}-p+\lambda(-p), \qquad \lambda\ge 0.
\end{align*}

Since $F>0$, $A(p)$ approaches infinity as $p$ approaches $\overline{p}$ from below. As $A(p)$ is continuous on $[0,\overline{p})$ and finite at $p=0$, a globally optimal price $p^*$ exists. At  $p^*$, the Kuhn--Tucker conditions imply that either $A'(p^*)=0$, or $p^*=0$ and $A'(0)>0$. In the second case, the non-negative pricing constraint binds, and increasing price from zero would increase average cost. Thus, a necessary condition for zero pricing to minimize average net expenditure per connection is:\begin{align*}
A'(0) = \frac{-F M'\left(0\right)}{M\left(0\right)^2}  - 1 = \frac {F}{\overline{M}} \eta\left(0\right) - 1 \geq 0, \qquad  \eta\left(0\right) =-\dfrac{M'(0)}{\overline{M}}
\end{align*}

\noindent where $\eta \left(0\right)$ defines the semi-elasticity of demand with respect to price at $p=0$, measuring the proportional reduction in take-up per unit increase in price from zero.

In other words, for zero pricing to minimize average cost, two conditions must be met: i) the fixed cost contribution to average cost ($\sfrac{F}{\overline{M}}$) is large; and ii) demand is highly sensitive to introducing even small prices. The second condition is documented in a wide variety of cases throughout the developing world \parencite{D2014,BGG2012,KM2007,CD2010,ABS2010,S2014}. How often the first condition is met is more uncertain, though the evidence suggests it most likely holds for networked infrastructure \parencite{LMW2020}.

These conditions are necessary but not sufficient. A higher price could decrease average cost relative to free provision even when imposing a small price increases average cost. The demand curve that we fit to our data illustrates the point: for a Weibull function with a shape parameter less than 1 (\weibullBPartRound{} in our data), the semi-elasticity of demand grows without bound as the price approaches zero, so the necessary condition for demand sensitivity at zero price is always met, and zero always yields a local minimum in average cost. But as the exercise in Section \ref{cost_eff} illustrates, zero may not yield a global minimum.

The result that the cost-effectiveness-maximizing price can be zero is driven by the presence of economies of scale, captured by the fixed costs $F$. Without fixed costs, setting $p=c$ would allow the policymaker to recover all costs, running the programme at zero net cost. This is consistent with the marginal-cost pricing principle established in the classical literature on welfare-maximizing infrastructure pricing \parencite[e.g.,][]{H1938}. However, setting $p=c$ would result in incomplete take-up, creating the standard equity-efficiency trade-off between cost recovery and universal access.

\section{Conclusion}\label{conclusion}

A key question for policymakers is how to price infrastructure connections. Setting prices too low forgoes the opportunity to raise revenues to offset costs, but setting prices too high reduces take-up and may exclude the poor, an apparent equity-efficiency trade-off. We conduct a willingness-to-pay elicitation to measure demand for household connection to piped drinking water, where truthful revelation of preferences is incentivized by the opportunity to connect to a real pipeline network that we construct. We combine this with project cost data to derive the cost-effectiveness-maximizing price.

We show that economies of scale in infrastructure provision strongly dominate the potential to raise revenues from user fees, such that the price that maximizes cost-effectiveness is zero. The conclusion is robust to analysis choices. Holding the shape of the demand curve fixed, willingness to pay would need to increase by a factor of at least \kScaleCEPositivePart{} for a positive price to improve cost-effectiveness relative to free provision. Our analysis also abstracts from the significant overhead costs necessary to collect and administer payments, which further reduce cost-effectiveness \parencite{CGH2026,R2009b}.  Even without considering these additional costs, economies of scale that are sufficiently strong remove the equity-efficiency trade-off in pricing decisions.

The importance of economies of scale is illustrated by the contrast between our results and those of \textcite{BJS2026}, who in the absence of any obvious economies of scale in safe water delivery describe a clear equity-efficiency trade-off: private provision could be self-sustaining at high prices, but only a few households would gain access to safe drinking water at these prices. For pricing decisions without economies of scale, a zero price will not generally maximize cost-effectiveness. This could include the pricing of water use once a network has been built. Building the network has significant economies of scale; operating it may not.

Although our results suggest that free provision of at least a basic level of service is likely to be a feature of any cost-effectiveness-maximizing strategy, alternative, more sophisticated pricing strategies might be even more cost-effective. For example, one could provide a basic level of service for free with optional, paid-for upgrades.  We saw some evidence of demand for these types of upgrades in the field. In our case, take-up of the piped water connection was 100\% when offered for free. With a good with public benefits and private costs, such as sewer connections, a negative price---an incentive paid to connectors---could increase cost-effectiveness over a zero price, as for vaccinations in \textcite{BDG2010}.

Our empirical analysis is limited to the case of piped water networks in rural Bangladesh, and our sample size is relatively small, but we show that a zero price can maximize cost-effectiveness when two conditions hold: high sensitivity of demand to prices near zero and significant economies of scale. The first condition is well documented to hold for poor households throughout the developing world \parencite[e.g.][]{BGG2012,D2014}. Whether rural pipeline networks exhibit unusually high economies of scale remains an open question. Systematic evidence on the existence of economies of scale in public service provision is scanty, but what evidence there is suggests moderately increasing to strong returns to scale, at least for networked infrastructure \parencite{G-RL-PM-V2023,LMW2020,CER2016}.

Our conclusion may be unsatisfying in that it leaves unresolved the problem of how policymakers should fund infrastructure with important economies of scale in poor communities. But the absence of an easy answer to this question does not make user fees the solution. In our setting, connection fees do not stretch a limited budget further; with a fixed budget, connection fees just result in fewer households being connected at higher average costs than could be achieved with free provision. Our central recommendation is that policymakers should carefully consider the possibility of economies of scale when they evaluate how to set user fees for infrastructure.

\clearpage

\small
\singlespacing
\linespread{1.0}

\printbibliography

\normalsize
\onehalfspacing

\singlespacing
\appendix

\newpage

	\titlespacing*{\section}{-20pt}{50pt}{20pt}
	\setcounter{tocdepth}{4}
	\setcounter{secnumdepth}{4}

\section*{Appendices}\label{appendix}

\vspace*{-0.5cm}
\setcounter{table}{0}
\setcounter{figure}{0}

\renewcommand{\thesubsection}{\Alph{subsection}}
\renewcommand{\thesubsubsection}{\thesubsection\arabic{subsubsection}}

\renewcommand{\thetable}{\Alph{subsection}\arabic{table}}
\renewcommand{\thefigure}{\Alph{subsection}\arabic{figure}}

\subsection{Appendix Figures and Tables}
\label{app:figs_tabs}

\begin{figure}[th]
\caption{Water tower for pipeline network\label{fig:water_tower}}
\begin{center}
\vspace{-2ex} \includegraphics[width=0.5\textwidth]{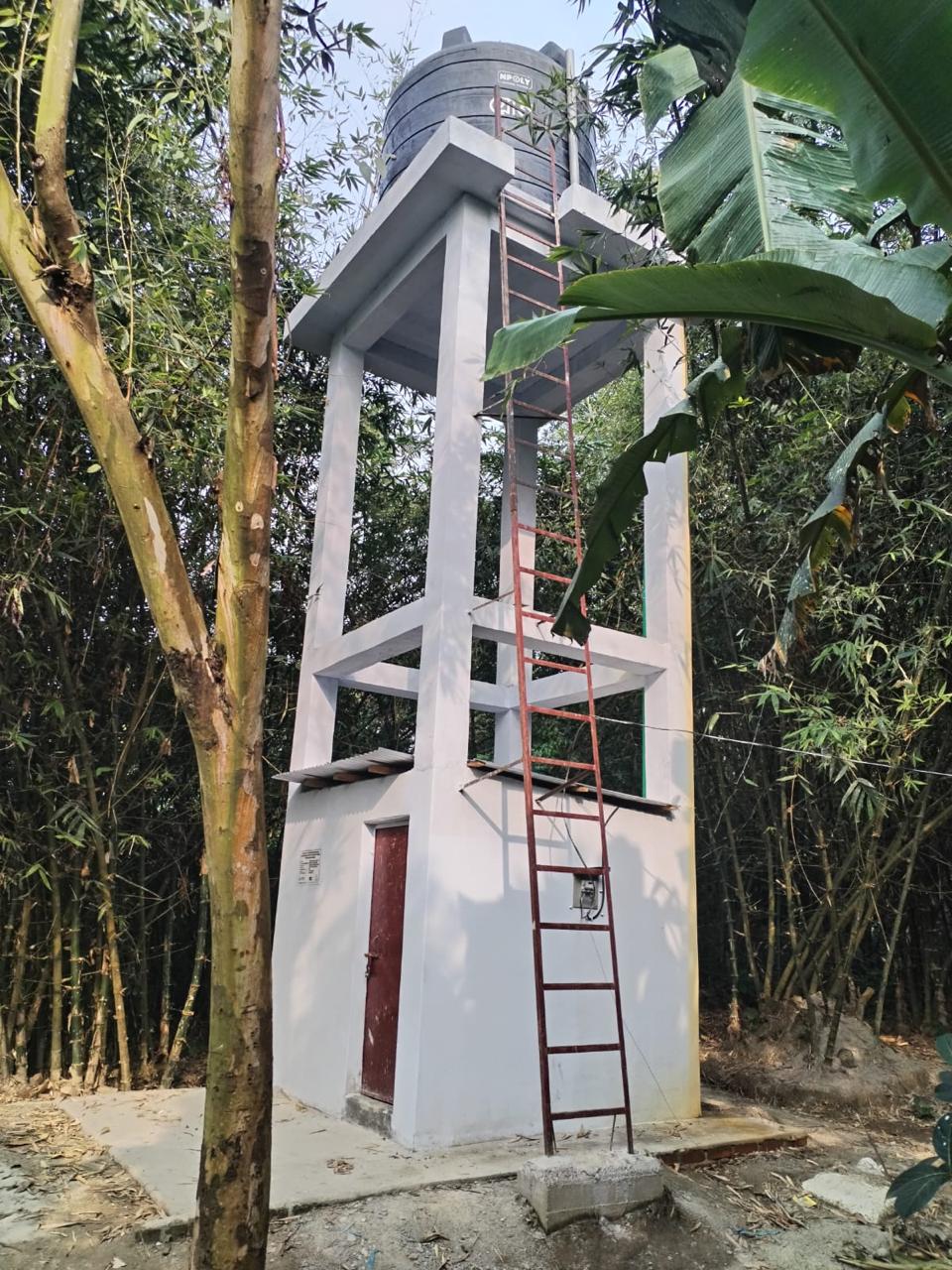}\\
\end{center}
{\vspace{-2ex} \footnotesize \emph{Notes} Photo credit: Md. Ahasan Habib.  \par}
\end{figure}

\begin{figure}[h]
\caption{Advantages of a pipeline connection\label{fig:advantages}}
\begin{center}
\vspace{-2ex} \includegraphics[width=\textwidth]{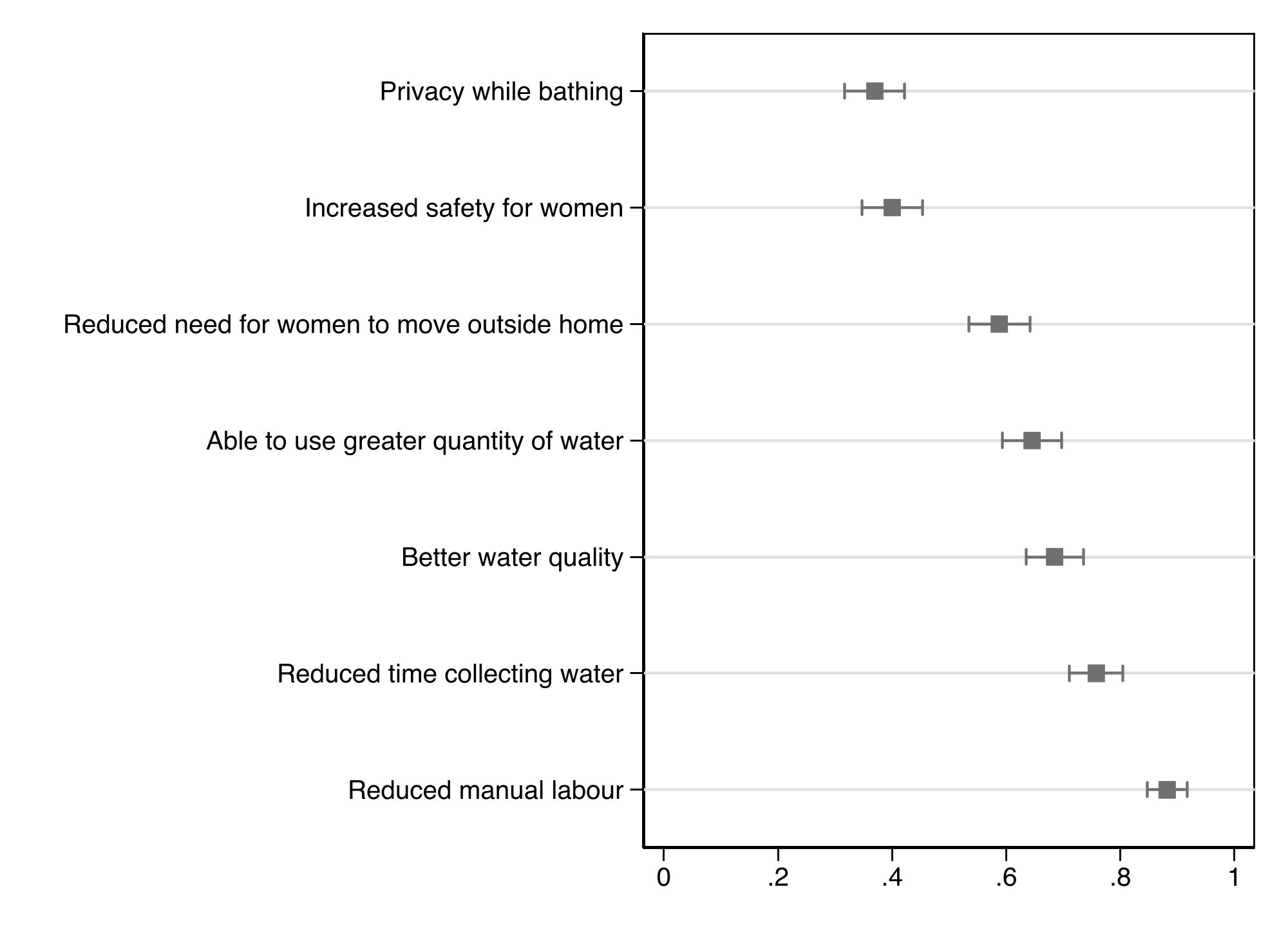}\\
\end{center}
{\vspace{-2ex} \footnotesize \emph{Notes} Graph shows share of respondents giving listed advantage when asked about the relative advantages of a household piped water connection. Sample includes only households interested in a piped water connection in study communities (\nInterested/\nEligible).  90\% confidence intervals reflect robust standard errors \parencite{LE2000}. \par}
\end{figure}

\begin{figure}[h]
\caption{Piped water price vouchers: drawn vs expected under the design\label{fig:voucher_design}}
\begin{center}
\vspace{-2ex} \includegraphics[width=0.8\textwidth]{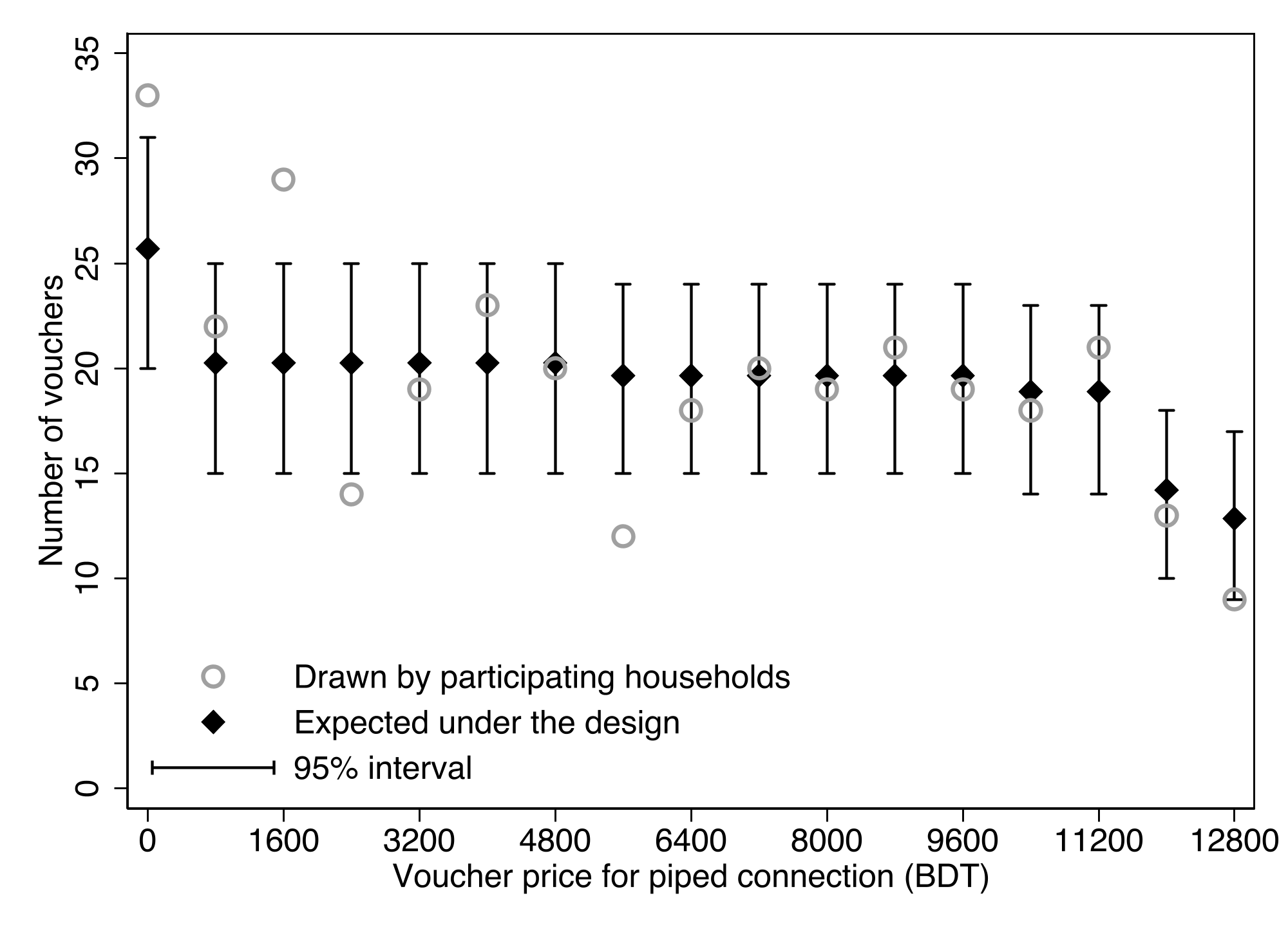}\\
\end{center}
{\vspace{-2ex} \footnotesize \emph{Notes} Graph shows the number of piped water price vouchers drawn at each price in Bangladeshi taka (BDT) by the \nWTPPart{} households who participated in the WTP elicitation (hollow circles), and the number expected under the voucher design given the number of participating households in each community (diamonds). Each household was shown five envelopes containing vouchers drawn at random from the pool and selected one; the other four were returned to the pool. Each community had its own pool of vouchers, which was larger than the community size given that each respondent needed to be presented with five vouchers. Whiskers show 95\% intervals for the number drawn at each price, assuming that each household drew one voucher at random, without replacement, from the pool in its community. Intervals are computed separately for each price; with 17 prices, one or two would be expected to fall outside their intervals by chance. The deviations from perfectly random allocation are most likely explained by imperfect shuffling of vouchers.  \par}
\end{figure}

\begin{figure}[h]
\caption{Voucher lottery randomization check: soap price vs assets \label{fig:price_wealth_soap}}
\begin{center}
\vspace{-2ex} \includegraphics[width=0.6\textwidth]{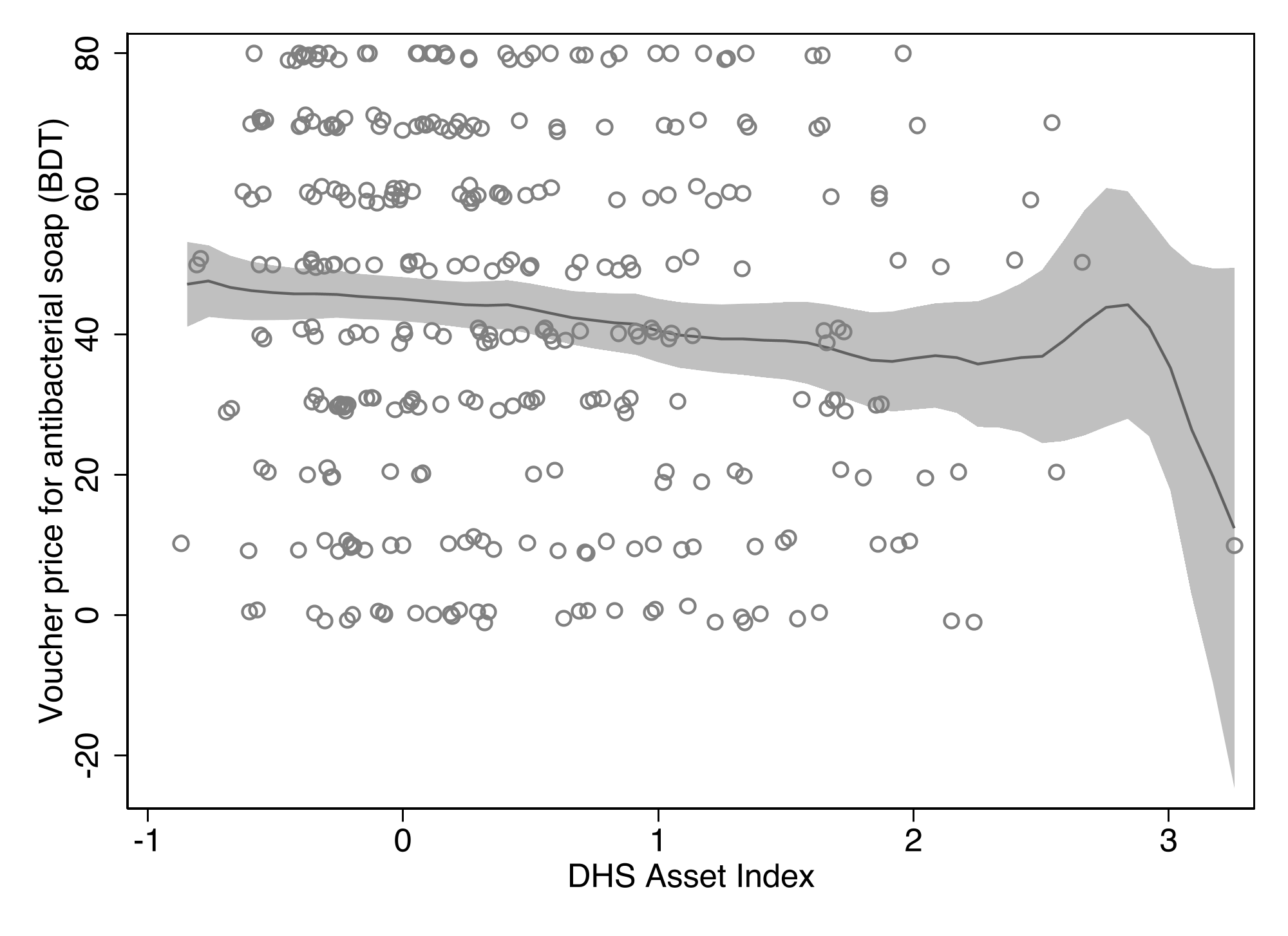}\\
\end{center}
{\vspace{-2ex} \footnotesize \emph{Notes} Graph shows randomly drawn price in Bangladeshi taka (BDT) for a bar of antibacterial soap plotted against a household-level asset index constructed using weights from the DHS surveys. Individual observations shown as hollow circles, jittered for readability. Line and shaded interval show the non-parametrically estimated relationship between the voucher price and the asset index with 90\% confidence interval. If the price lottery occurred as designed, we should observe no correlation between the two. Note the sparsity of observations at very high wealth levels and the corresponding width of the confidence interval. \par}
\end{figure}

\begin{figure}[h]
\caption{Voucher lottery randomization check: piped water price vs assets \label{fig:price_wealth}}
\begin{center}
\vspace{-2ex} \includegraphics[width=0.6\textwidth]{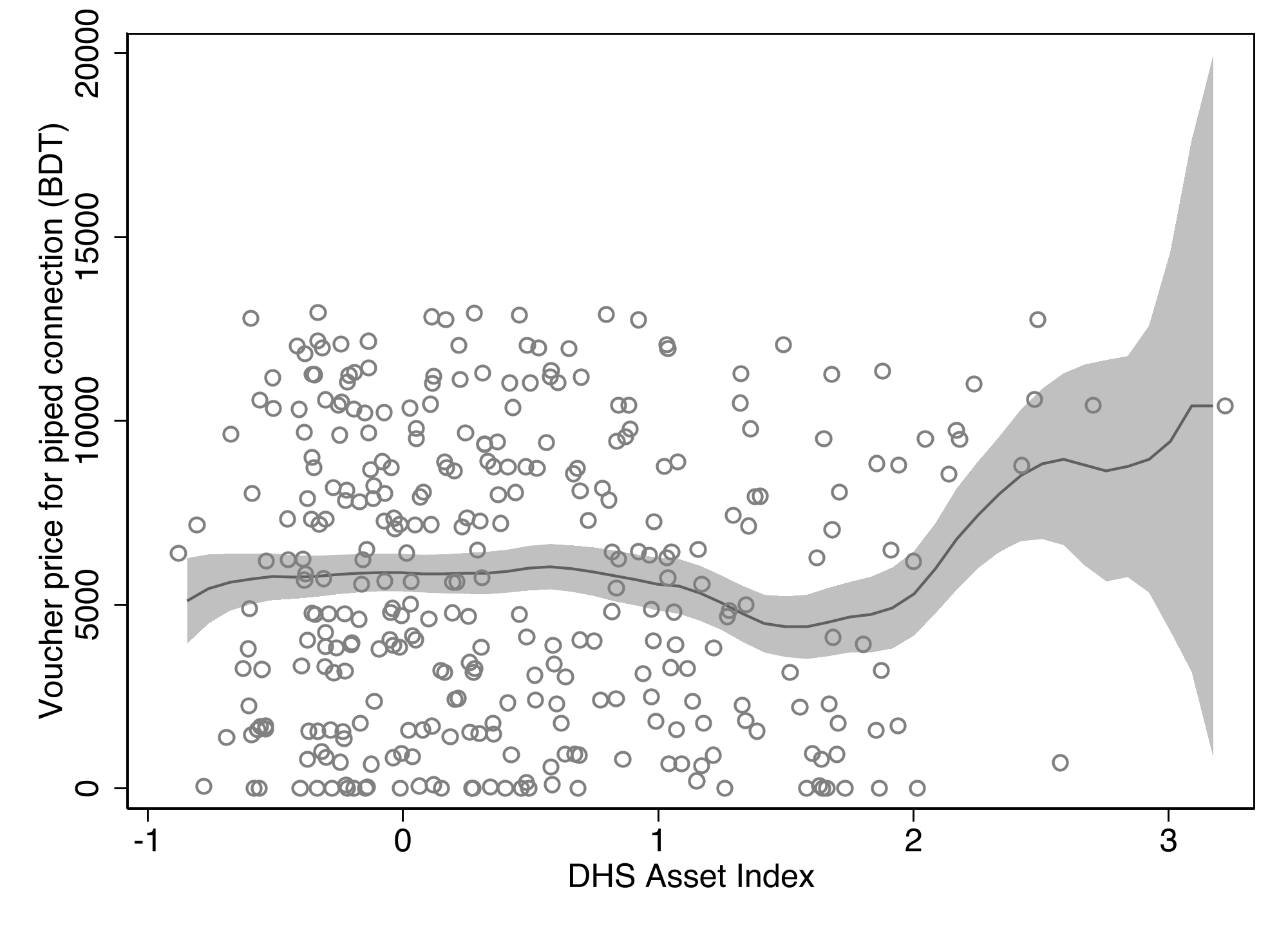}\\
\end{center}
{\vspace{-2ex} \footnotesize \emph{Notes} Graph shows randomly drawn price in Bangladeshi taka (BDT) for a piped water connection plotted against a household-level asset index constructed using weights from the DHS surveys. Individual observations shown as hollow circles, jittered for readability. Line and shaded interval show the non-parametrically estimated relationship between the voucher price and the asset index with 90\% confidence interval. If the price lottery occurred as designed, we should observe no correlation between the two. Note the sparsity of observations at very high wealth levels and the corresponding width of the confidence interval. \par}
\end{figure}

\begin{figure}[h]
\caption{Voucher lottery randomization check: soap price vs piped water price \label{fig:price_soap_pipe}}
\begin{center}
\vspace{-2ex} \includegraphics[width=0.6\textwidth]{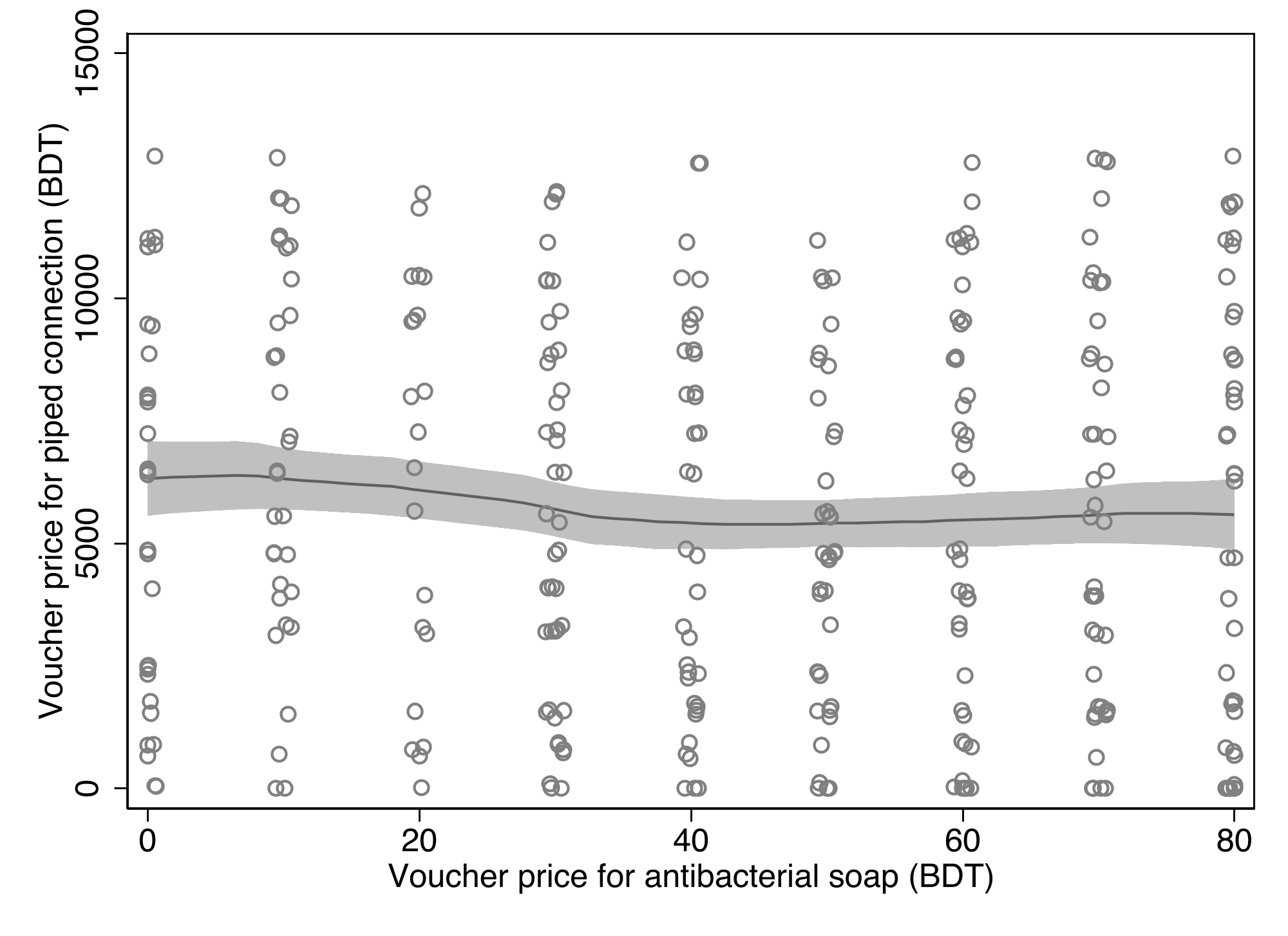}\\
\end{center}
{\vspace{-2ex} \footnotesize \emph{Notes} Graph shows randomly drawn price in Bangladeshi taka (BDT)  for a piped water connection against randomly drawn price for a bar of antibacterial soap. Individual observations shown as hollow circles, jittered for readability. Line and shaded interval show the non-parametrically estimated relationship between the two prices with 90\% confidence interval and bandwidth set to 10. If the price lottery occurred as designed, we should observe no correlation between the two.  \par}
\end{figure}

\begin{figure}[h]
\caption{Share of households reporting distress\label{fig:WTP_distress}}
\begin{center}
\vspace{-2ex} \includegraphics[width=0.6\textwidth]{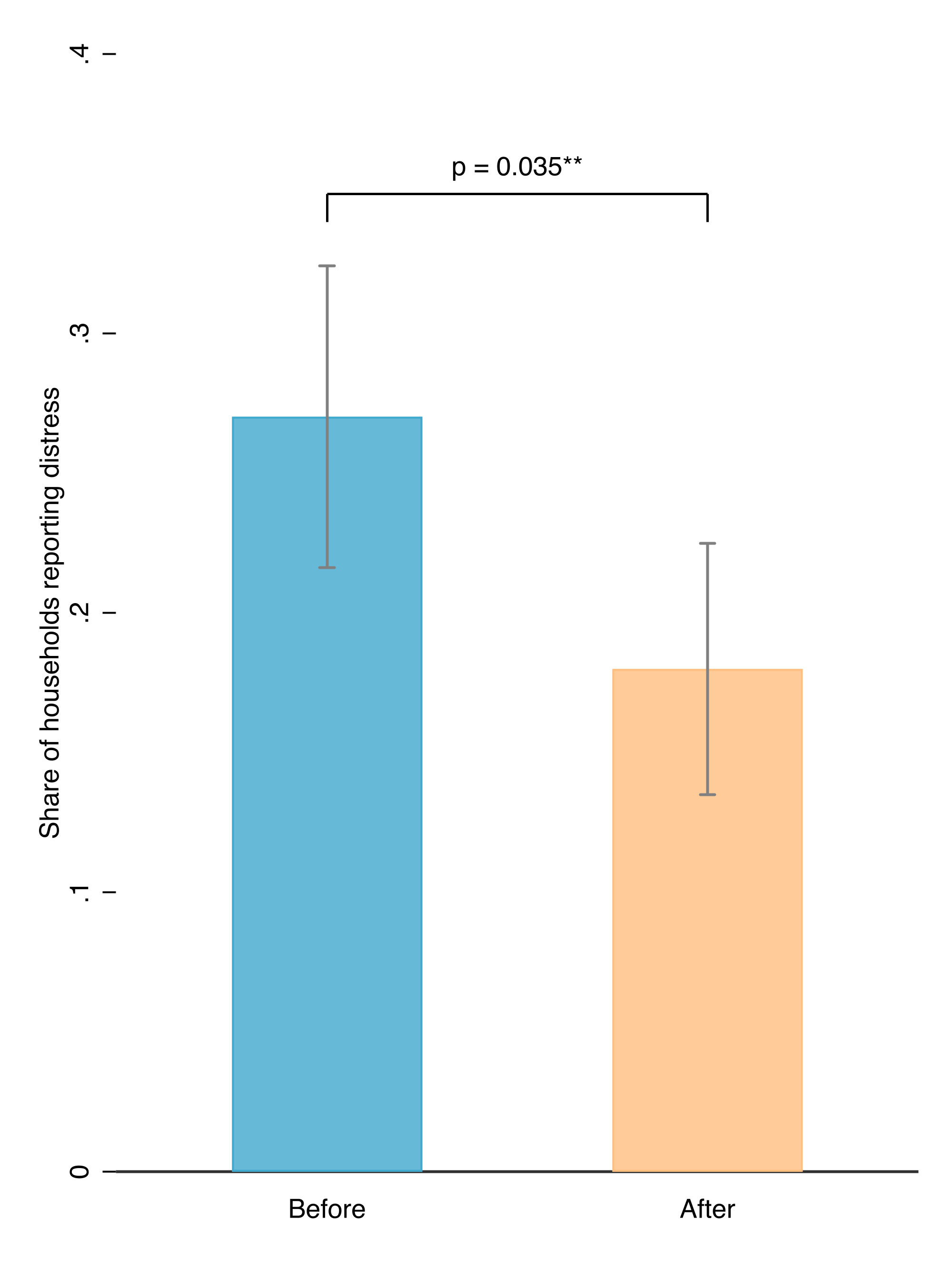}\\
\end{center}
{\vspace{-2ex} \footnotesize \emph{Notes} Graph shows share of respondents replying in the affirmative when asked whether anyone in the household was currently experiencing a high level of stress or any emotional distress. Households randomly assigned to be asked the question immediately before or after the willingness-to-pay elicitation.  Question was asked to all households who consented to participate in the survey, independent of their decision to participate in the WTP elicitation.  90\% confidence intervals reflect robust standard errors \parencite{LE2000}. \par}
\end{figure}

\begin{figure}[h]
\caption{Willingness-to-pay comparison\label{fig:pipe_vs_soap}}
\begin{center}
\vspace{-2ex} \includegraphics[width=\textwidth]{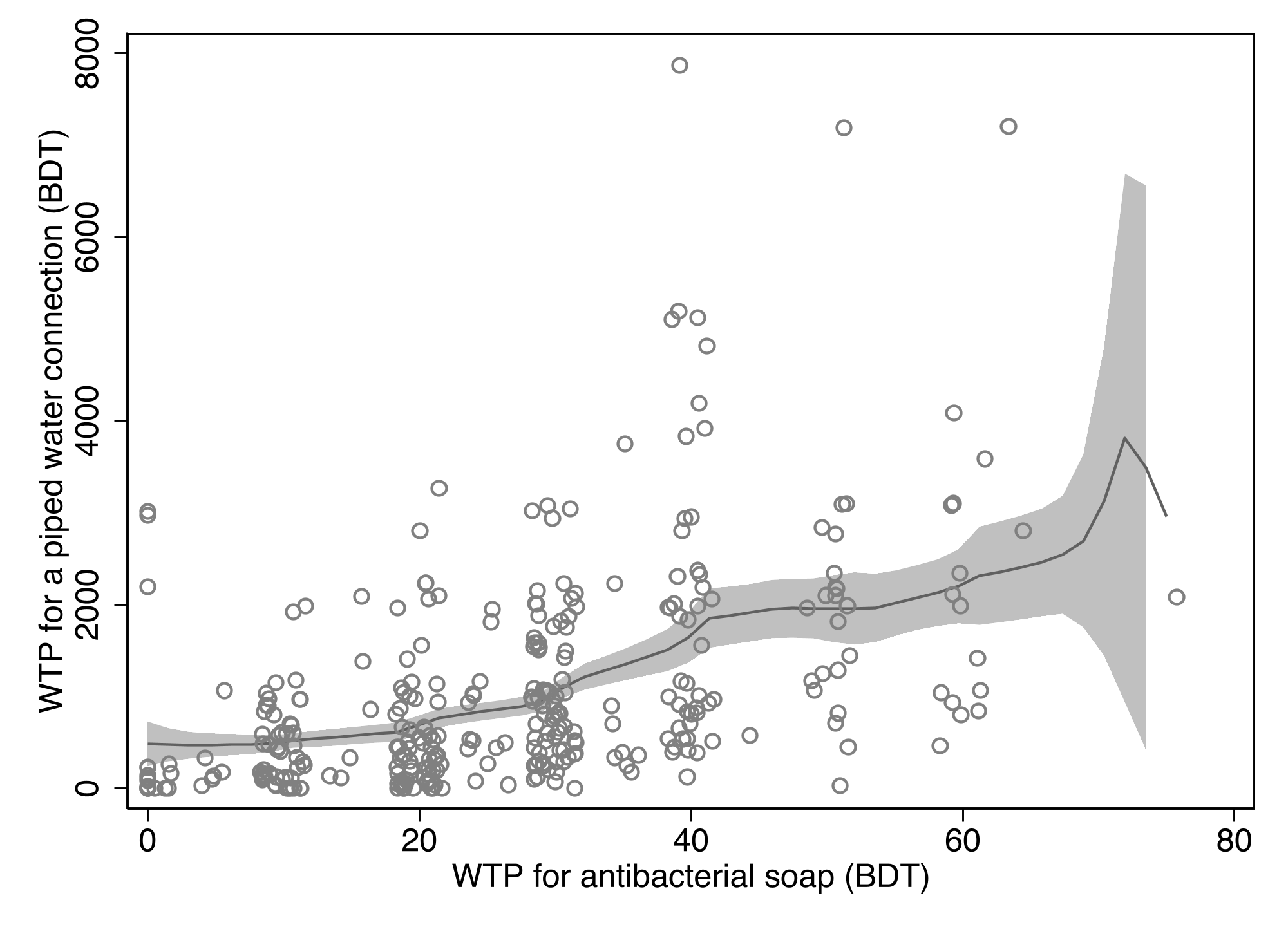}\\
\end{center}
{\vspace{-2ex} \footnotesize \emph{Notes} Graph shows willingness to pay (WTP) in Bangladeshi taka (BDT) for a bar of antibacterial soap and a piped water connection to the household.  Both WTP elicitations were incentivized. Individual observations shown as hollow circles, jittered for readability. Line and shaded interval show the non-parametrically estimated relationship between the two WTP valuations with 90\% confidence interval. \par}
\end{figure}

\begin{figure}[h]
\caption{Demand and average revenue: assigning WTP of zero to households that declined to participate\label{fig:rev_max_usd_all}}
\begin{center}
\vspace{-2ex} \includegraphics[width=0.8\textwidth]{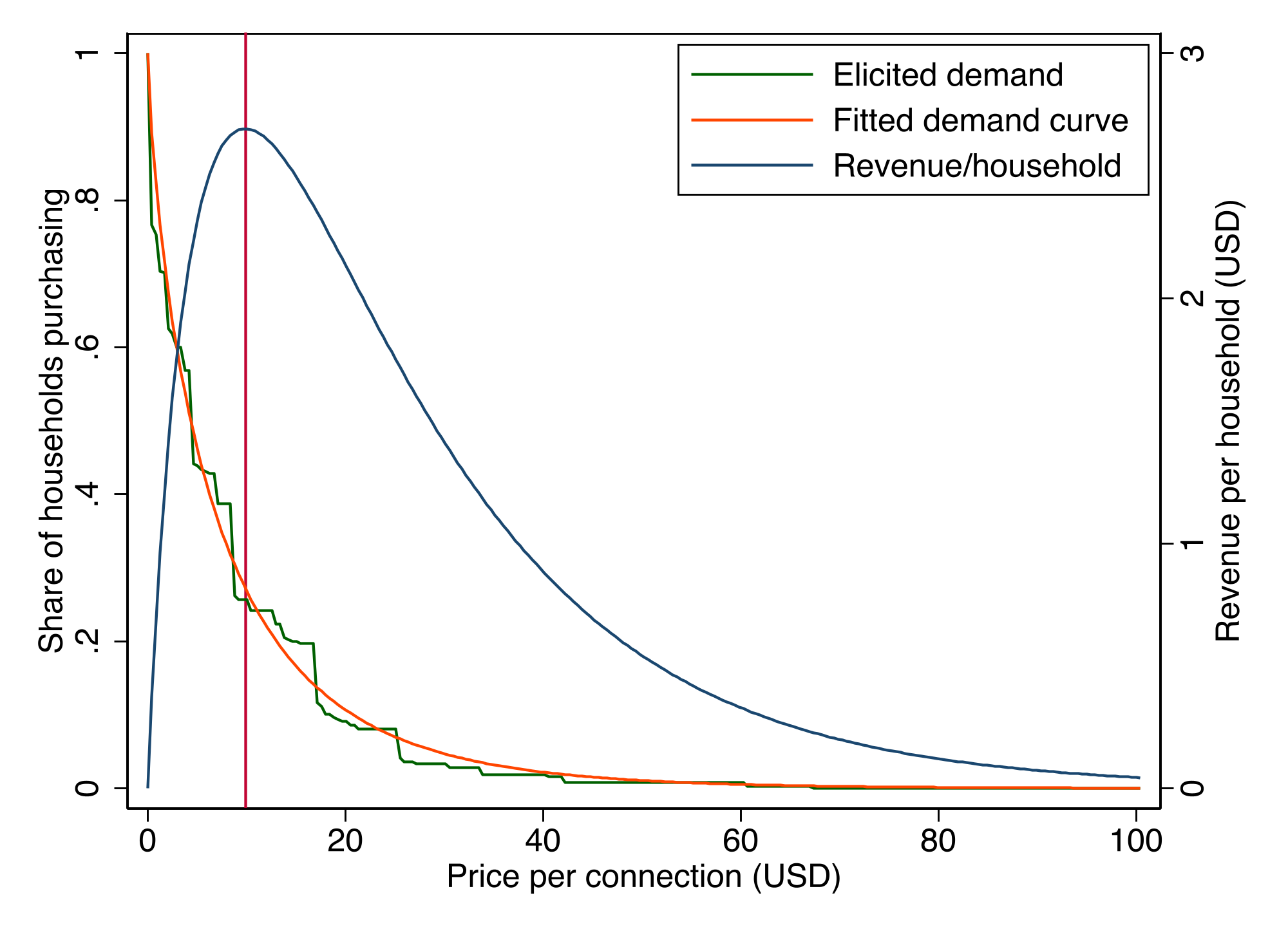}\\
\end{center}
{\vspace{-2ex} \footnotesize \emph{Notes} $N$ = \nWTPSample{} eligible households that participated in the household survey. Households that declined to participate in the willingness-to-pay elicitation are assigned a WTP of zero. Raw data from willingness-to-pay elicitation shown in green.  Weibull function fitted to elicited demand data shown in orange ($e^{-\weibullAUSD \, p^{\weibullBRound}}$, where $p$ is the price in US dollars). Revenues per household shown in blue. \par}
\end{figure}

\begin{figure}[h]
\caption{Demand by community: participating households\label{fig:demand_by_village}}
\begin{center}
\vspace{-2ex} \includegraphics[width=0.8\textwidth]{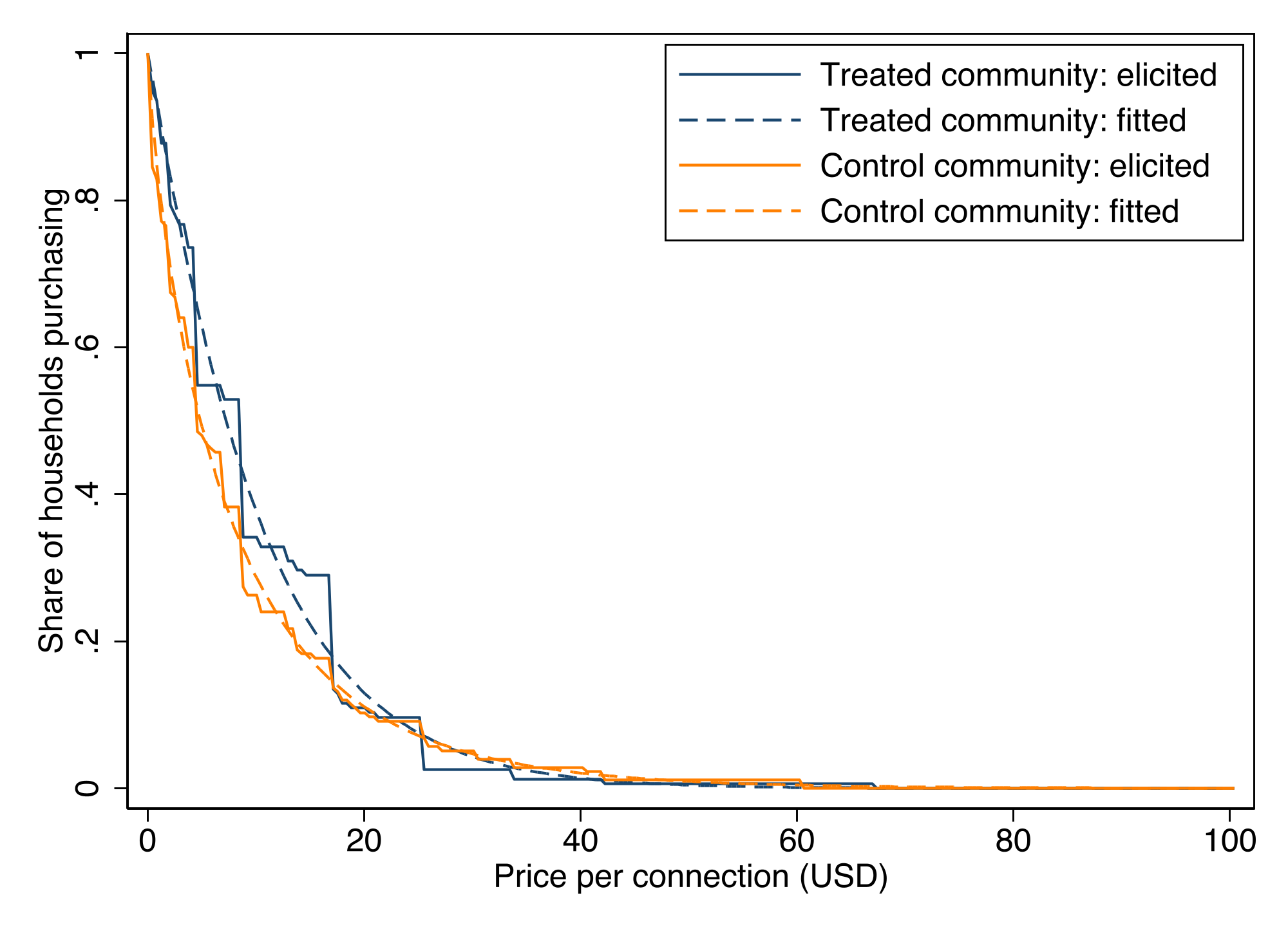}\\
\end{center}
{\vspace{-2ex} \footnotesize \emph{Notes} Eligible households that participated in the willingness-to-pay elicitation, by community. Solid lines show raw data from the elicitation; dashed lines show Weibull functions fitted separately for each community. \par}
\end{figure}

\begin{figure}[h]
\caption{Demand by community: assigning WTP of zero to households that declined to participate\label{fig:demand_by_village_all}}
\begin{center}
\vspace{-2ex} \includegraphics[width=0.8\textwidth]{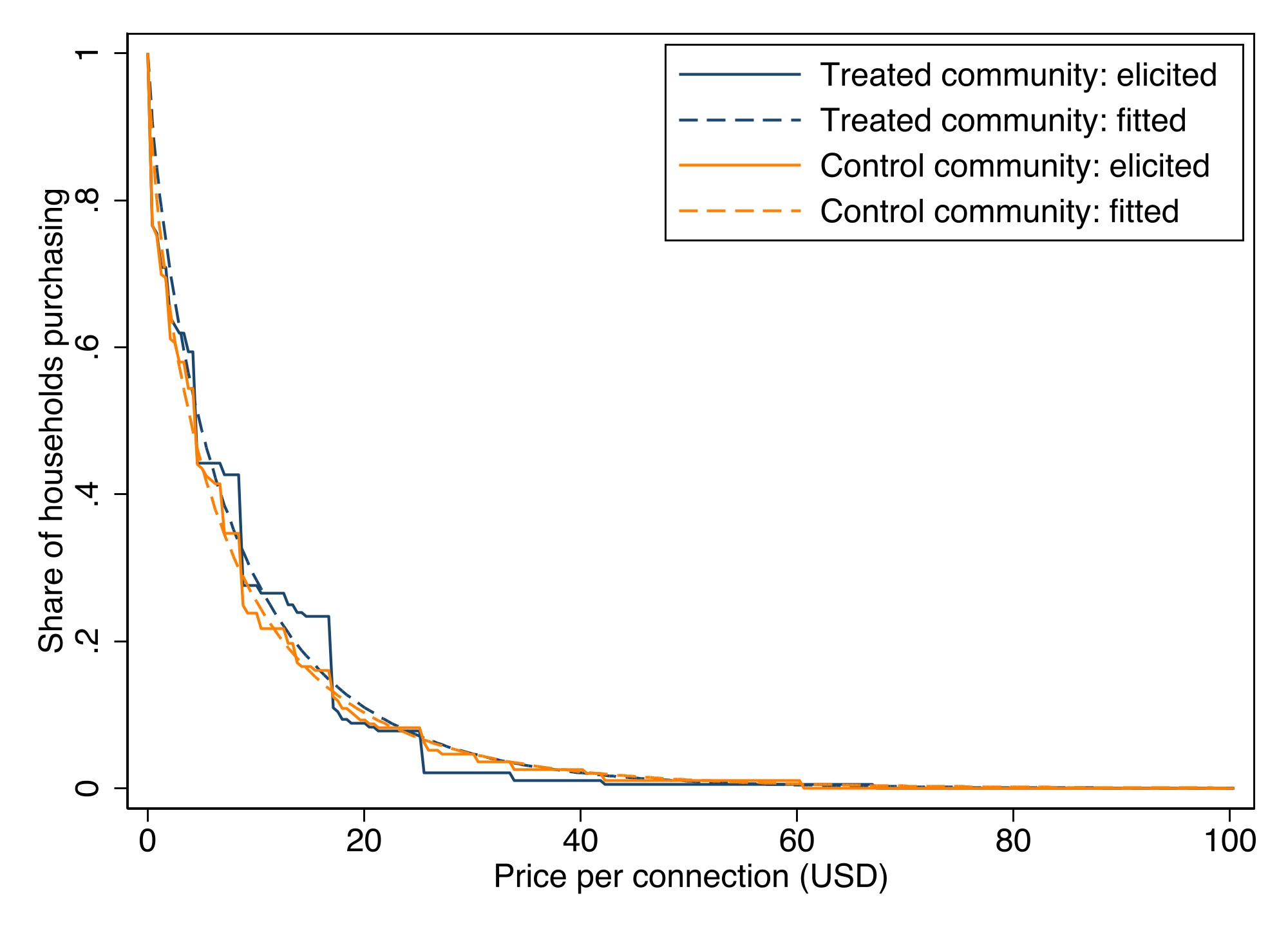}\\
\end{center}
{\vspace{-2ex} \footnotesize \emph{Notes} Eligible households that participated in the household survey, by community. Households that declined to participate in the willingness-to-pay elicitation are assigned a WTP of zero. Solid lines show raw data from the elicitation; dashed lines show Weibull functions fitted separately for each community. \par}
\end{figure}

\begin{figure}[h]
\caption{Willingness to pay: estimated vs expected \label{fig:obs_vs_exp}}
\begin{center}
\vspace{-2ex} \includegraphics[width=\textwidth]{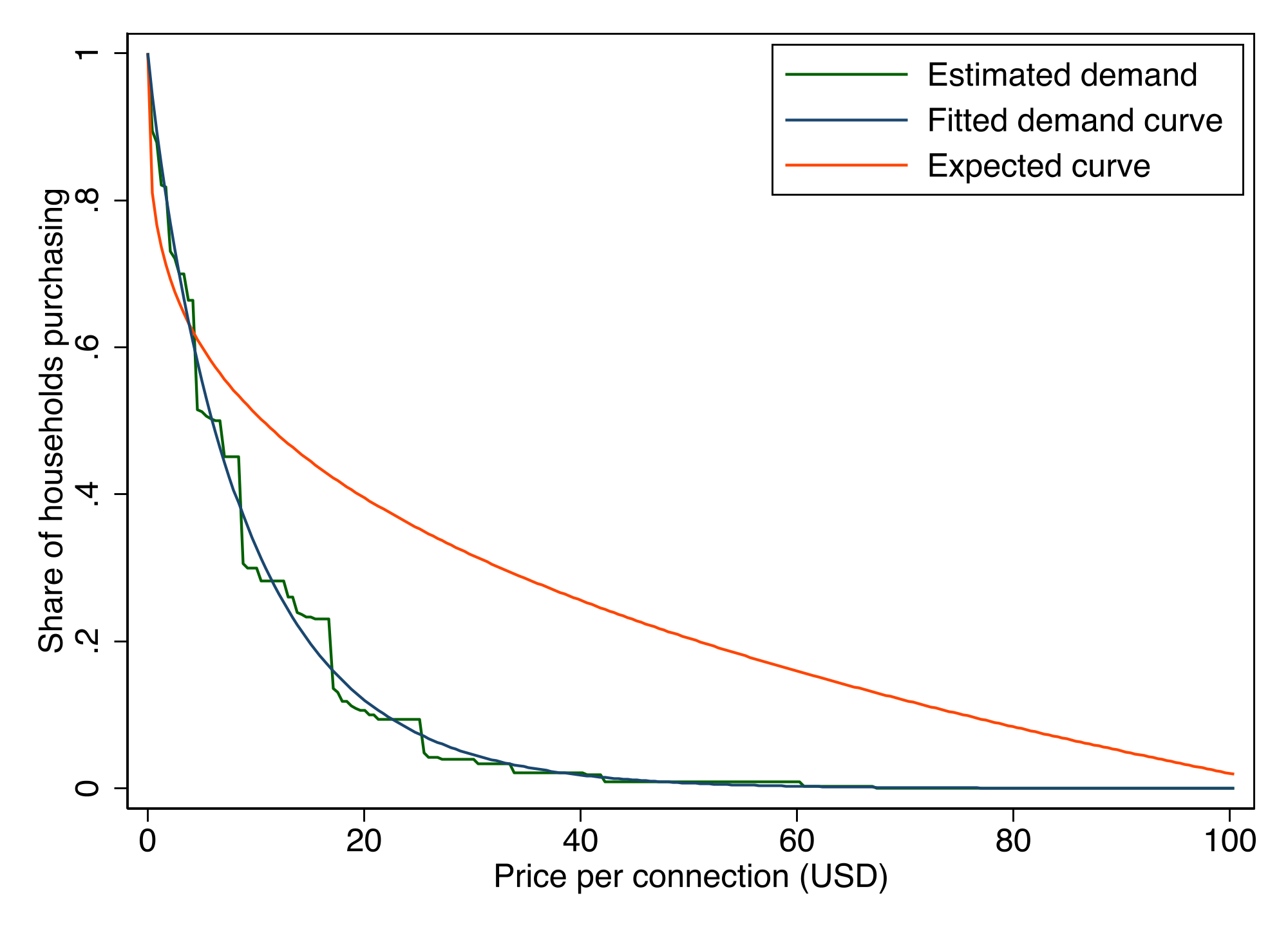}\\
\end{center}
{\vspace{-2ex} \footnotesize \emph{Notes} $N$ = \nWTPPart{} eligible households that participated in the willingness-to-pay elicitation. Raw data from willingness-to-pay elicitation shown in green.  Weibull function fitted to elicited demand data shown in blue ($e^{-\weibullAPartUSD \, p^{\weibullBPartRound}}$, where $p$ is the price in US dollars). Ex ante expected demand curve used to design the price lottery shown in orange. \par}
\end{figure}

\begin{figure}[h]
\caption{Schematic map of arsenic contamination in treated community\label{fig:arsenic_map}}
\begin{center}
\vspace{-2ex} \includegraphics[height=0.55\textheight]{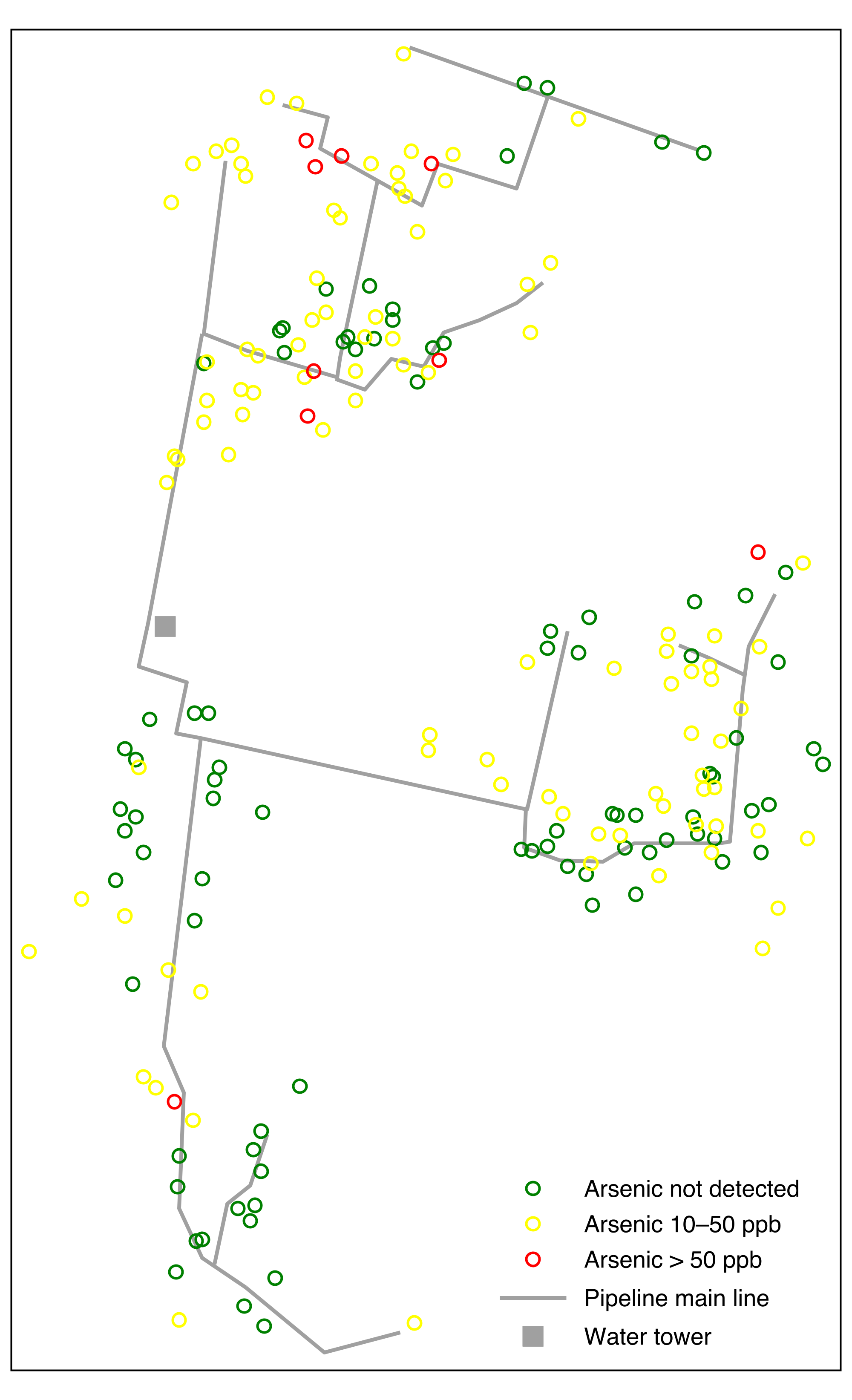}\\
\end{center}
{\vspace{-2ex} \footnotesize \emph{Notes} Household location data based on GPS coordinates collected in 2023. The grey line shows the main line of the piped water network and the grey square the water tower. Background imagery and geographical identifiers are not shown to preserve the anonymity of the study community.    \par}
\end{figure}

\begin{figure}[h]
\caption{Reasons for not purchasing a pipeline connection\label{fig:reasons_no_connect}}
\begin{center}
\vspace{-2ex} \includegraphics[width=\textwidth]{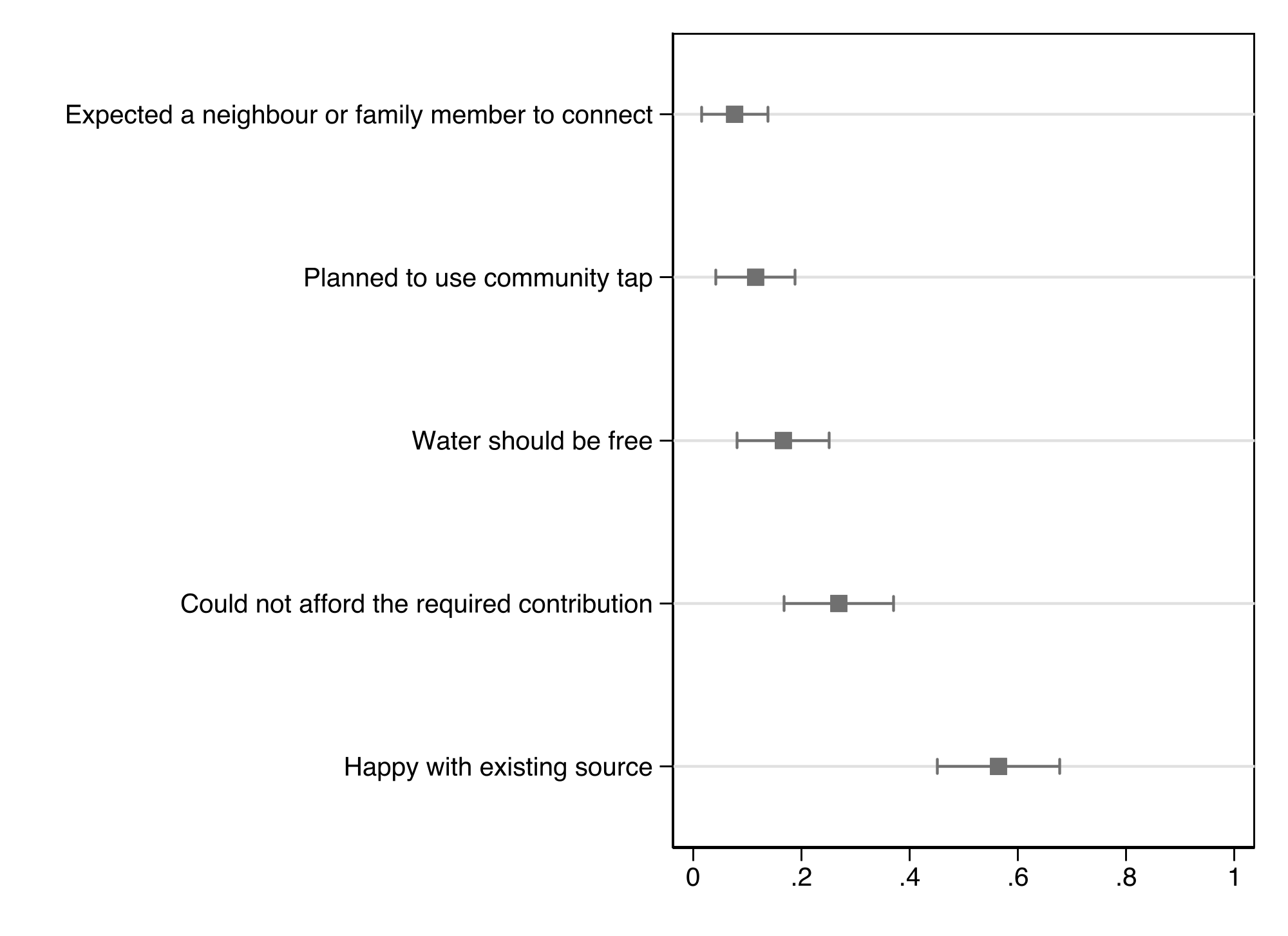}\\
\end{center}
{\vspace{-2ex} \footnotesize \emph{Notes} Graph shows share of respondents giving listed reason when asked in the 2024 household survey why they decided not to purchase a piped water connection. Respondents could give more than one reason. Sample includes only eligible households in the treated community that did not connect to the pipeline according to project records and that answered the question (\nNoConnectReasons{} households). Reasons given by fewer than 5\% of respondents are not shown. 90\% confidence intervals reflect robust standard errors \parencite{LE2000}. \par}
\end{figure}

\begin{figure}[h]
\caption{Arsenic contamination of wells in 2015 and 2023\label{fig:arsenic_persistence}}
\begin{center}
\vspace{-2ex} \includegraphics[width=0.7\textwidth]{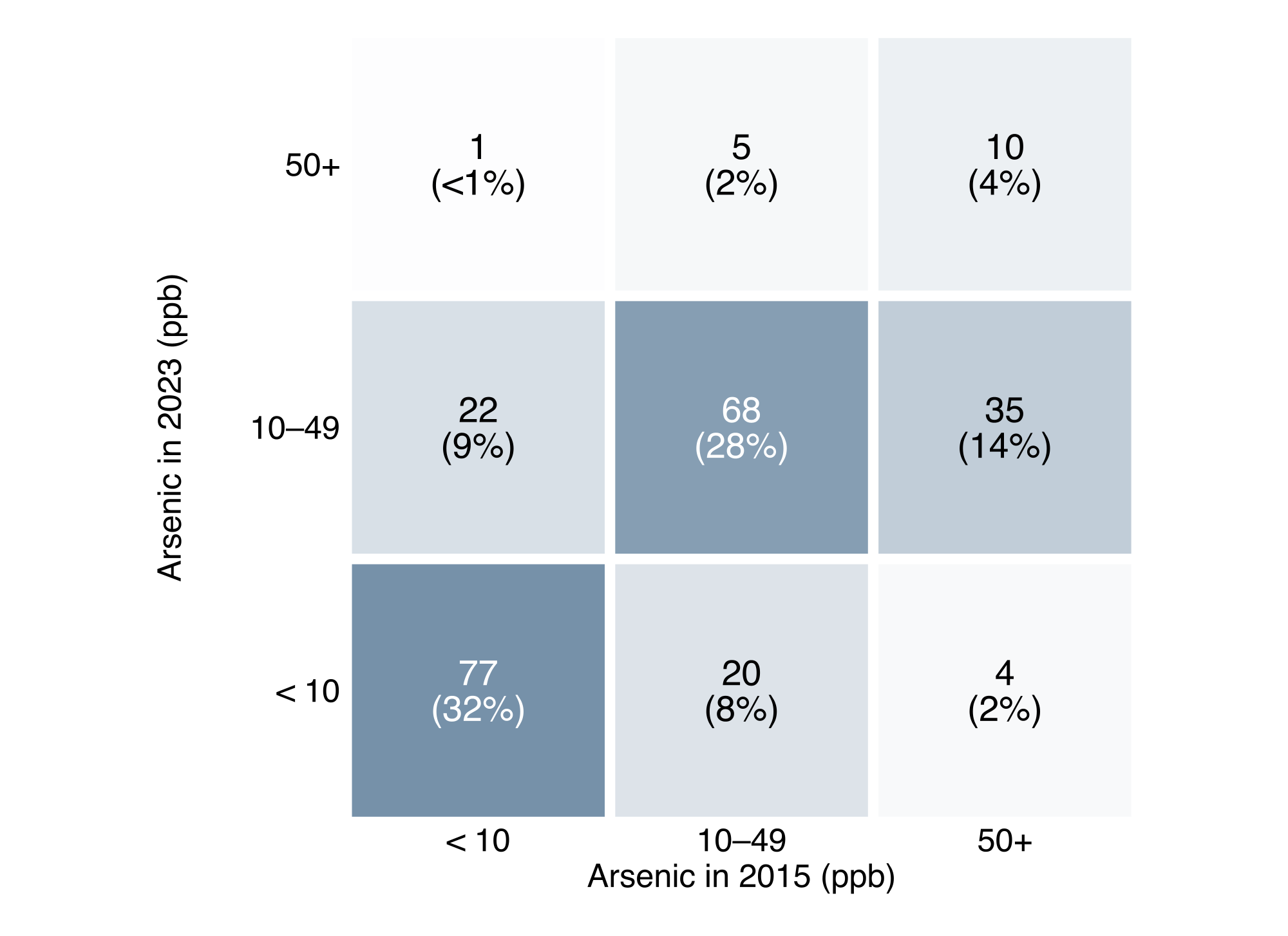}\\
\end{center}
{\vspace{-2ex} \footnotesize \emph{Notes} Wells in the two study communities tested for arsenic in both 2015 and 2023 (\nWellsBothTests{} wells), by arsenic contamination in each year relative to the WHO and Bangladeshi exposure thresholds: no detectable arsenic (below 10~ppb), 10--49~ppb, and 50~ppb or more. Each cell shows the number of wells and, in parentheses, the share of all wells; darker shading indicates a larger share. Cells on the diagonal contain wells in the same category in both years. The 2015 and 2023 tests used different field test kits. \par}
\end{figure}

\begin{figure}[h]
\caption{Demand curves with higher willingness to pay\label{fig:wtp_scale_demand}}
\begin{center}
\vspace{-2ex} \includegraphics[width=0.95\textwidth]{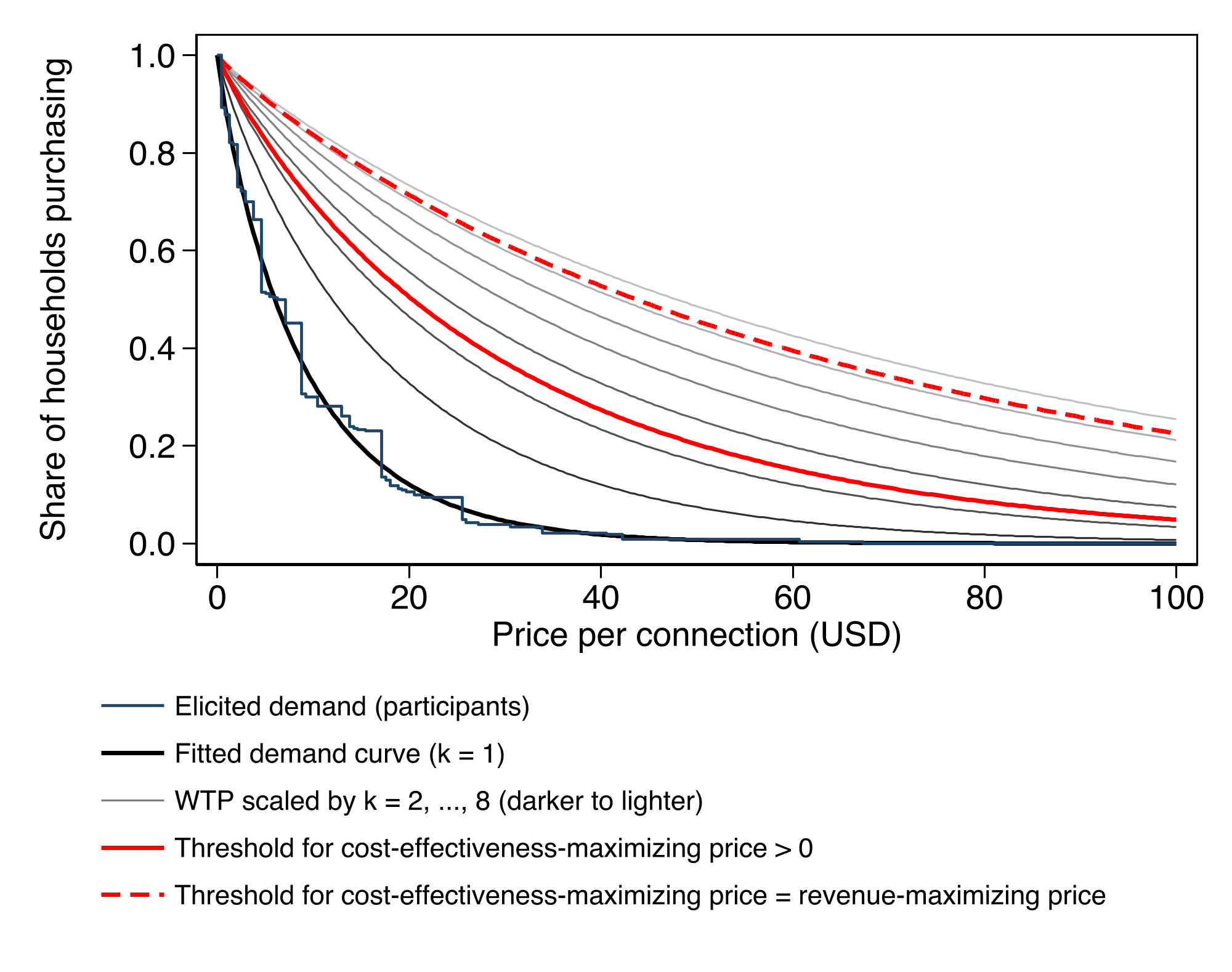}\\
\end{center}
{\vspace{-2ex} \footnotesize \emph{Notes} Demand for a household connection when the willingness to pay of every household that participated in the WTP elicitation is multiplied by a scale factor $k$. Navy: demand revealed in the WTP elicitation (participating households, both communities). Black: Weibull demand curve fitted to elicited demand ($k = 1$). Grey: $k = 2, \ldots, 8$ (darker to lighter), the fitted curve with the same shape parameter and willingness to pay multiplied by $k$. Solid red: $k = \kScaleCEPositivePart$, the scale factor above which the cost-effectiveness-maximizing price exceeds zero. Dashed red: $k = \kScaleCEEqualsRevPart$, the scale factor at which the cost-effectiveness-maximizing price equals the revenue-maximizing price (Figure \ref{fig:wtp_scale}). \par}
\end{figure}

\clearpage

\begin{table}[h]
\caption{Example of the dynamic price list\label{tab:dynamic_price_list}}
\begin{center}
\begin{tabular}{cccc}
\toprule
Question & Price asked (BDT) & Answer & Interval for WTP after answer (BDT) \\
\midrule
1 & 6,400 & No  & [0, 6,400) \\
2 & 3,200 & No  & [0, 3,200) \\
3 & 1,600 & No  & [0, 1,600) \\
4 & 800   & Yes & [800, 1,600) \\
5 & 1,200 & No  & [800, 1,200) \\
6 & 1,000 & Yes & [1,000, 1,200) \\
7 & 1,100 & No  & [1,000, 1,100) \\
8 & 1,050 & No  & [1,000, 1,050) \\
\midrule
\multicolumn{4}{l}{Recorded WTP: 1,000 BDT} \\
\bottomrule
\end{tabular}
\end{center}
{\vspace{-1ex} \footnotesize \emph{Notes} Example sequence of questions asked to a household willing to pay between 1,000 and 1,050 BDT for a connection. Each question asks ``Would you pay [price] taka for a household water connection?''. The first price asked is always 6,400 BDT, the midpoint of the range of possible voucher prices (0 to 12,800 BDT). After each answer, the next price asked is the midpoint of the remaining interval. Each response narrows the interval: a ``yes'' raises the lower bound to the price asked, and a ``no'' lowers the upper bound to the price asked. After eight questions, the interval is 50 BDT (US\$\wtpIntervalUSD) wide, and the recorded WTP is its lower bound, the highest price the household agreed to pay. The enumerator confirmed the recorded WTP with the respondent, revising the answers if the respondent disagreed, before the respondent drew a price voucher. \par}
\end{table}

\begin{table}[h]
\caption{Observed take-up and take-up predicted by the WTP elicitation\label{tab:takeup_prediction}}
\begin{center}
\small
\input{tables/takeup_prediction.tex}
\end{center}
{\footnotesize \emph{Notes} Eligible households in the treated community, excluding households that drew a voucher for a free connection in the price lottery. Each column reports a regression of observed take-up at a price of 1000 BDT on an indicator for predicted take-up, equal to one if the household's WTP in the elicitation was 1000 BDT or more. Column 1: households that participated in the WTP elicitation. Column 2: all households that consented to the survey, assigning a WTP of zero to households that declined to participate in the WTP elicitation. Households that participated in the WTP elicitation are reweighted by the inverse of the probability of drawing a non-zero voucher, so that the sample in column 2 remains representative of eligible households in the community; in column 1, all households receive the same weight. Robust standard errors in parentheses \parencite{LE2000}. \par}
\end{table}

\begin{table}[h]
\caption{Effects of arsenic on take-up\label{tab:arsenic_takeup}}
\begin{center}
\small
\input{tables/arsenic_takeup.tex}
\end{center}
{\footnotesize \emph{Notes} Treated community, except column 5, which pools both communities. Each column reports a regression of take-up at a price of 1000 BDT on three indicators for the arsenic contamination of the household's water source at baseline, without a constant; coefficients are therefore group means. Observed take-up: the household connected to the pipeline network. Predicted take-up: the household reported a WTP of at least 1000 BDT in the elicitation. Columns 1--3 exclude households that drew a free connection in the price lottery; columns 4 and 5 include them. Column 1 includes households that declined to participate in the WTP elicitation and reweights the remaining participants by the inverse of the probability of drawing a non-zero price. Columns 2--5 include only households that participated in the WTP elicitation. Robust standard errors in parentheses \parencite{LE2000}. \par}
\end{table}

\begin{table}[h]
\caption{Reported safety of water source by measured arsenic contamination\label{tab:beliefs_arsenic}}
\begin{center}
\small
\input{tables/beliefs_pooled.tex}
\end{center}
{\footnotesize \emph{Notes} Households that participated in the household survey and whose water source is a tubewell. Column 1: the household reported at baseline that the source had been tested for arsenic. Column 2: the household reported that the source had been tested and that the test showed it to be safe (zero if the household reported that the source had not been tested or did not know). Column 3: the household reported that the test showed the source to be safe, among households that reported that the source had been tested (the question was asked only of these households). Column 4: the spout of the source was painted green, indicating that it had been tested and found safe. Each column reports a regression on three indicators for the arsenic contamination of the source measured at baseline, without a constant; coefficients are therefore group means. Robust standard errors in parentheses \parencite{LE2000}. \par}
\end{table}

\begin{landscape}
\begin{table}[p]
\caption{Reported safety of water source by arsenic contamination measured in 2015\label{tab:beliefs_arsenic_2015}}
\begin{center}
\small
\input{tables/beliefs_2015.tex}
\end{center}
{\footnotesize \emph{Notes} Column 1: all wells in the two study communities tested for arsenic in 2015; the outcome is an indicator for the well being used for drinking or cooking by at least one surveyed household in 2023. Columns 2--5: households that participated in the household survey and whose water source is a tubewell tested for arsenic in 2015 (\nHHSourceTestedCCLPG{} households); outcomes are defined as in Appendix Table \ref{tab:beliefs_arsenic}. Each column reports a regression on three indicators for the arsenic contamination of the source measured in 2015, without a constant; coefficients are therefore group means. Robust standard errors in parentheses \parencite{LE2000}. \par}
\end{table}
\end{landscape}

\end{document}

%% file: numbers.tex
\newcommand{\nEligible}{422}
\newcommand{\nSurvey}{385}
\newcommand{\nWTPDeclined}{55}
\newcommand{\nCensusHH}{456}
\newcommand{\nCensusMigrated}{8}
\newcommand{\nCensusMerged}{3}
\newcommand{\nCensusFound}{445}
\newcommand{\nOutsideArea}{23}
\newcommand{\pctSurveyEligible}{91}
\newcommand{\nListedHH}{488}
\newcommand{\nListedFound}{439}
\newcommand{\nListedMigrated}{27}
\newcommand{\nListedNoSurvivors}{17}
\newcommand{\nListedMislisted}{5}
\newcommand{\nCensusNew}{17}
\newcommand{\nHeadChanged}{24}
\newcommand{\nResurvey}{376}
\newcommand{\pctResurvey}{97.7}
\newcommand{\nResurveyNoSurvivors}{3}
\newcommand{\nResurveyDeclined}{1}
\newcommand{\nResurveyNotFound}{5}
\newcommand{\nDeclinedAll}{37}

\newcommand{\nVoucherPrices}{17}
\newcommand{\voucherStepUSD}{7}
\newcommand{\voucherMaxUSD}{107}
\newcommand{\wtpIntervalUSD}{0.42}

\newcommand{\nSoapBelowWTP}{118}
\newcommand{\pctSoapBought}{98}
\newcommand{\nInterested}{329}
\newcommand{\pctInterested}{85}
\newcommand{\meanAdvantages}{4.4}

\newcommand{\pctPredTakeupThousandPart}{53}

\newcommand{\nFreeVoucherTreated}{20}

\newcommand{\nVoucherRetained}{33}
\newcommand{\nVoucherEightHundredRetained}{6}
\newcommand{\nVoucherEightHundredRetainedConn}{2}
\newcommand{\nTreatedSample}{192}
\newcommand{\nTreatedSampleConnected}{113}
\newcommand{\pctObsTakeupThousand}{54}
\newcommand{\pctObsTakeupThousandPart}{55}

\newcommand{\pctTakeupLowWTPPart}{53}
\newcommand{\pctTakeupHighWTPPart}{56}
\newcommand{\pTakeupWTPPart}{0.761}
\newcommand{\pctTakeupNonPart}{51}
\newcommand{\pctTakeupNoArsenic}{40}
\newcommand{\pctTakeupArsenicTen}{64}
\newcommand{\pctTakeupArsenicFifty}{86}
\newcommand{\nMapHH}{192}
\newcommand{\nNoConnectReasons}{78}
\newcommand{\nWTPPart}{330}
\newcommand{\pctWTPZeroPart}{11}
\newcommand{\wtpMedianUSDPart}{6.3}
\newcommand{\wtpNinetyUSDPart}{21}
\newcommand{\wtpMaxUSDPart}{67}
\newcommand{\wtpMeanUSDPart}{9}
\newcommand{\costPerHHIdealUSDPart}{89}

\newcommand{\pctIdealSavingPart}{9}

\newcommand{\weibullBPartRound}{0.92}
\newcommand{\weibullAPartUSD}{0.133}
\newcommand{\revMaxPriceUSDPart}{9.7}

\newcommand{\pctTakeupOneUSDPart}{88}

\newcommand{\pctTakeupTwentyUSDPart}{12}
\newcommand{\kScaleCEPositivePart}{3.4}
\newcommand{\kScaleCEEqualsRevPart}{7.3}

\newcommand{\nWTPSample}{385}
\newcommand{\pctWTPDeclined}{14}

\newcommand{\weibullBRound}{0.77}
\newcommand{\weibullAUSD}{0.223}

\newcommand{\fixedCostUSD}{4,795}
\newcommand{\varCostUSD}{74}
\newcommand{\capacityHH}{200}
\newcommand{\nPipelineLaid}{176}
\newcommand{\totalCostUSD}{17,819}
\newcommand{\costPerLaidUSD}{101}

\newcommand{\runCostMonthUSD}{44.8}

\newcommand{\nConnectedTotal}{135}
\newcommand{\runCostMonthPerHHUSD}{0.33}

\newcommand{\costPerHHFreeUSD}{98}

\newcommand{\capPriceUSD}{8.4}
\newcommand{\pctWTPDeclinedTreat}{19}
\newcommand{\pctWTPDeclinedCtrl}{9}

\newcommand{\ksPVillagePart}{0.052}

\newcommand{\ksPVillageAll}{0.535}

\newcommand{\pctSafeIfTested}{94}
\newcommand{\pctPaintedGreen}{8}

\newcommand{\nWellsListedCensus}{350}

\newcommand{\nWellsCensusGone}{46}
\newcommand{\nWellsCensusNew}{84}
\newcommand{\nSourcesUsedBase}{355}
\newcommand{\nSourcesUnusedEnd}{57}
\newcommand{\nConnected}{109}
\newcommand{\nHHNewTubewell}{13}
\newcommand{\nNewTubewellsTreated}{2}
\newcommand{\nNewTubewellsControl}{11}

\newcommand{\nWellsBothTests}{242}

\newcommand{\nHHSourceTestedCCLPG}{263}

%% file: tables/summary_demand.tex
\begin{tabular}{l>{\centering\arraybackslash}p{1.6cm}>{\centering\arraybackslash}p{1.9cm}>{\centering\arraybackslash}p{2.2cm}>{\centering\arraybackslash}p{2.2cm}>{\centering\arraybackslash}p{3.4cm}>{\centering\arraybackslash}p{0.8cm}}
\toprule
 & (1) & (2) & (3) & (4) & (5) & (6) \\
 & Median WTP (US\$) & Take-up at 1000 BDT (US\$8.4) & Revenue-maximizing price (US\$) & Take-up at revenue-maximizing price & Cost-effectiveness-maximizing price (US\$) & N \\
\midrule
\multicolumn{7}{l}{\emph{Panel A: WTP elicitation}} \\
Participated in WTP elicitation &  6.3 &  45\% &  9.7 &  34\% &   0 & 330 \\
All eligible (WTP = 0 if declined) &  4.2 &  39\% &  9.9 &  27\% &   0 & 385 \\
Treated, participants &  8.4 &  53\% &  9.7 &  39\% &   0 & 155 \\
Treated, all eligible (WTP = 0 if declined) &  4.2 &  43\% &  9.8 &  29\% &   0 & 192 \\
Control, participants &  4.2 &  38\% &  9.8 &  29\% &   0 & 175 \\
Control, all eligible (WTP = 0 if declined) &  4.2 &  35\% & 10.0 &  26\% &   0 & 193 \\
\midrule
\multicolumn{7}{l}{\emph{Panel B: Observed take-up (treated community)}} \\
Participated in WTP elicitation & &  55\% & & & & 135 \\
All eligible & &  54\% & & & & 172 \\
\bottomrule
\end{tabular}

%% file: tables/takeup_prediction.tex
\begin{tabular}{l*{2}{>{\centering\arraybackslash}p{2.6cm}}}
\toprule
 & \multicolumn{2}{c}{Observed take-up at 1000 BDT} \\
 & (1) & (2) \\
\midrule
Predicted take-up (WTP $\geq$ 1000 BDT) & 0.03 & 0.03 \\
 & (0.09) & (0.08) \\
Constant & 0.53 & 0.53 \\
 & (0.07) & (0.05) \\
\midrule
Mean predicted take-up & 0.56 & 0.45 \\
Mean observed take-up & 0.55 & 0.54 \\
N & 135 & 172 \\
\addlinespace
Includes non-participants, assigning WTP = 0 & No & Yes \\
\bottomrule
\end{tabular}

%% file: tables/arsenic_takeup.tex
\begin{tabular}{l*{5}{>{\centering\arraybackslash}p{1.8cm}}}
\toprule
 & \multicolumn{2}{c}{Observed take-up} & \multicolumn{3}{c}{Predicted take-up} \\
 & (1) & (2) & (3) & (4) & (5) \\
\midrule
No arsenic ($<$ 10 ppb) & 0.40 & 0.42 & 0.55 & 0.54 & 0.53 \\
 & (0.06) & (0.06) & (0.06) & (0.06) & (0.05) \\
Arsenic 10--50 ppb & 0.64 & 0.64 & 0.59 & 0.55 & 0.41 \\
 & (0.05) & (0.06) & (0.06) & (0.06) & (0.04) \\
Arsenic $>$ 50 ppb & 0.86 & 0.86 & 0.29 & 0.33 & 0.36 \\
 & (0.15) & (0.15) & (0.20) & (0.18) & (0.09) \\
\midrule
$p$ value: equal across groups & 0.001 & 0.007 & 0.340 & 0.518 & 0.089 \\
N & 171 & 135 & 135 & 155 & 329 \\
\addlinespace
Sample & Treated & Treated & Treated & Treated & Both \\
Includes non-participants & Yes & No & No & No & No \\
Includes zero price vouchers & No & No & No & Yes & Yes \\
\bottomrule
\end{tabular}

%% file: tables/beliefs_pooled.tex
\begin{tabular}{l*{4}{>{\centering\arraybackslash}p{2.3cm}}}
\toprule
 & (1) & (2) & (3) & (4) \\
 & Reports source tested & Reports source tested and safe & Reports source safe, if tested & Spout painted green \\
\midrule
No arsenic ($<$ 10 ppb) & 0.69 & 0.68 & 0.98 & 0.11 \\
 & (0.04) & (0.04) & (0.01) & (0.02) \\
Arsenic 10--50 ppb & 0.72 & 0.66 & 0.91 & 0.06 \\
 & (0.03) & (0.03) & (0.02) & (0.02) \\
Arsenic $>$ 50 ppb & 0.71 & 0.64 & 0.90 & 0.04 \\
 & (0.09) & (0.09) & (0.07) & (0.04) \\
\midrule
Sample mean & 0.71 & 0.66 & 0.94 & 0.08 \\
$p$ value: equal across groups & 0.789 & 0.927 & 0.032 & 0.147 \\
N & 379 & 379 & 268 & 379 \\
\bottomrule
\end{tabular}

%% file: tables/beliefs_2015.tex
\begin{tabular}{l*{5}{>{\centering\arraybackslash}p{3.0cm}}}
\toprule
 & (1) & (2) & (3) & (4) & (5) \\
 & Source still in use in 2023 & Reports source tested & Reports source tested and safe & Reports source safe, if tested & Spout painted green \\
\midrule
No arsenic in 2015 ($<$ 10 ppb) & 0.65 & 0.67 & 0.64 & 0.94 & 0.12 \\
 & (0.04) & (0.05) & (0.05) & (0.03) & (0.03) \\
Arsenic 10--50 ppb in 2015 & 0.73 & 0.80 & 0.76 & 0.95 & 0.07 \\
 & (0.04) & (0.04) & (0.04) & (0.02) & (0.03) \\
Arsenic $>$ 50 ppb in 2015 & 0.75 & 0.85 & 0.70 & 0.83 & 0.06 \\
 & (0.05) & (0.05) & (0.06) & (0.06) & (0.03) \\
\midrule
Sample mean & 0.70 & 0.76 & 0.70 & 0.92 & 0.09 \\
$p$ value: equal across groups & 0.239 & 0.020 & 0.126 & 0.126 & 0.290 \\
N & 349 & 263 & 263 & 200 & 263 \\
\bottomrule
\end{tabular}